\documentclass[aps,pra,twocolumn,superscriptaddress,showpacs]{revtex4-1}

\usepackage{amsfonts}
\usepackage{amsmath}
\usepackage{amssymb}

\usepackage{color}
\usepackage[dvipsnames]{xcolor}

\definecolor{darkblue}{rgb}{0.1,0.2,0.6}
\definecolor{darkred}{rgb}{0.8,0.1,0.2}
\definecolor{Gray}{gray}{0.9}
\usepackage[colorlinks,citecolor=darkblue,linkcolor=darkred,urlcolor=darkblue]{hyperref}	
\usepackage[all]{hypcap} %
\usepackage{graphicx}
\graphicspath{{images/}}
\usepackage{placeins}
\usepackage{floatrow}
\DeclareMathAlphabet{\mymathbb}{U}{BOONDOX-ds}{m}{n}
\usepackage{algorithm}
\usepackage{algpseudocode}
\usepackage{physics}
\newcommand{\avg}[1]{\left< #1 \right>}
\newcommand{\ie}{{\it i.e.~}}

\newcommand{\eg}{{\it e.g.~}}
\newcommand{\FF}{\mathbb{F}}

\begin{abstract}
  In this paper we present a novel method to generate hard instances with
  planted solutions based on the public-private McEliece post-quantum
  cryptographic protocol. Unlike other planting methods rooted in the
  infinite-size statistical analysis, our cryptographic protocol generates
  instances which are \emph{all} hard (in cryptographic terms), with the
  hardness tuned by the size of the private key, and with a guaranteed unique
  ground state. More importantly, because of the private-public key protocol,
  planted solutions cannot be easily recovered by a direct inspection of the
  planted instances without the knowledge of the private key used to generate
  them, therefore making our protocol suitable to test and evaluate quantum
  devices without the risk of ``backdoors'' being exploited.
\end{abstract}

\begin{document}

\def\quail{
	Quantum Artificial Intelligence Lab. (QuAIL), 
    NASA Ames Research Center, Moffett Field, CA 94035, USA
}
\def\kbr{
  KBR, Inc., 601 Jefferson St., Houston, TX 77002, USA
}

\title{Generating Hard Ising Instances With Planted Solutions Using Post-Quantum Cryptographic Protocols}
\author{Salvatore Mandr\`a}
\email{salvatore.mandra@nasa.gov}
\affiliation{\kbr}
\affiliation{\quail}

\author{Humberto Munoz-Bauza}
\affiliation{\kbr}
\affiliation{\quail}

\author{Gianni Mossi}
\affiliation{\kbr}
\affiliation{\quail}

\author{Eleanor G. Rieffel}
\affiliation{\quail}

\def\zo{\left\{0,\,1\right\}}

\maketitle

\section{Introduction}

With the advent of more competitive (either quantum \cite{arute2019quantum,
morvan2023phase, kim2023evidence}, quantum inspired \cite{kim2021physics,
aramon2019physics, zhu2015efficient} or post-Von Neumann
\cite{mohseni2022ising, sao2019application, traversa2015universal,
adleman1994molecular}) technologies for the optimization of classical cost
functions, it is becoming of utmost importance to identify classes of instances
which are well suited to transversally benchmark such devices. One of the major
challenges in designing planted problems is the generation of instances which
are hard to optimize: indeed, planting a desired solution may affect the energy
landscape of the instance, making it more convex and therefore easier to
explore. Another important aspect that may affect the hardness of planted
problems is the number of solutions: while it is guaranteed that the planted
configuration \emph{is} a solution, the planting protocol may introduce
multiple solutions that would make the instance easier to solve.

In this work we introduce a way of generating Ising instances with one unique
planted solution, whose hardness against optimization algorithms is derived
from the  robustness of the McEliece cryptographic protocol that is used to
create them. The rest of the paper is organized as follows: in Section
\ref{sec:planting_overview} we give an overview of the currently available
planting methods for Ising Hamiltonians, in Section \ref{sec:linear_codes} we
review linear codes, and the McEliece cryptographic protocol. Section
\ref{sec:hard_instances} shows how to generate hard instances of Ising
Hamiltonians with planted solutions by recasting the McEliece protocol in Ising
spin language, while Section \ref{sec:clustering} uses the machinery of
statistical physics to illustrate why energy landscapes analogous to the ones
exhibited by these hard instances should be difficult to explore for stochastic
local search algorithms. In Section \ref{sec:numerics}, we compare the
performance of Parallel Tempering on the Ising formulation of the McEliece
random instances against the best-known attack on the McEliece protocol.

\section{Overview of Existing Planting Methods}\label{sec:planting_overview}

For Constraint Satisfaction Problems (CSPs) like the graph coloring
\cite{papadimitriou1998combinatorial, sipser1996introduction} and the SAT
problem \cite{biere2009handbook}, solutions can be planted by using the
``quiet'' planting technique \cite{qu1999hiding, barthel2002hiding,
krzakala2009hiding, zdeborova2011quiet, krzakala2012reweighted,
sicuro2021planted}: constraints that satisfy the desired planted configuration
are added until the required density of constraints is achieved.  Similarly to
random CSPs \cite{mezard2002analytic, zdeborova2007phase}, planted CSPs have
different phase transitions by varying the density of constraints
\cite{zdeborova2007phase, krzakala2009hiding, ricci2019typology,
semerjian2020recovery}: while the ``satisfiability'' threshold is of course
suppressed, phase transitions related to the hardness of the instances, such as
the ``freezing'' transition \cite{zdeborova2007phase}, are preserved in planted
instances. In particular, in frozen planted instances, solutions are all
clustered into a single component which make the instance hard to optimize
\cite{zdeborova2008constraint}. While the quiet planting can be used to quickly
generate planted instances, properties like the hardness of the instances and
the number of solutions are only true for the ``typical'' case and in the limit
of a large number of variables, making the technique less valuable for
small-scale and mid-scale benchmarks.

Similar to the quiet planting, the ``patch'' planting is a technique used to
plant solutions for cost functions where not all the constraints must be
satisfied at the same time \cite{denchev2016computational, hen2015probing,
wang2017patch, hamze2018near, perera2020computational}. More precisely,
multiple subsystems with a known solution are ``patched'' together by properly
applying hard constraints between subsystems. Solutions for each subsystem are
found by either brute force or using other optimization techniques, and global
solutions are obtained by patching together solutions from all the subsystems.
The hardness of the patch planted instances is then tuned by increasing the
size and the hardness of the subsystems \cite{wang2017patch}.  Even if the
patch planting technique can generate instances rather hard to optimize
\cite{hamze2018near}, one of the main downsides of the technique is that a
direct inspection of the instances may easily reveal the solution. A random
permutation of the variable indexes could mitigate the problem, but the use of
cluster techniques \cite{zhu2015efficient} or the knowledge of the
pre-permutation layout may still be used to identify each subsystem.

Another common technique to plant solutions in combinatorial optimization
problems consists in using exactly solvable models that present phase
transitions. The simplest example is the problem of solving linear equations on
the binary field (also called XORSAT \cite{hen2019equation}). More precisely,
given a binary matrix $\mathbf{A}$ of size $M\times N$, with $M$ being the
number of equations and $N$ the number of variables, the XORSAT problem
consists in finding the solution $x$ such that $\mathbf{A} x = 0$, where the
identity is modulo 2. If $M > N$, the fundamental theorem of linear algebra
ensures that there are at least $2^{M-N}$ solutions. As for many other CSPs,
XORSAT instances undergo different phase transitions by changing the density of
constraints $M/N$, with the hardest instances being close to the satisfiability
threshold \cite{mezard2002analytic}. While XORSAT instances are known to be
hard to optimize using local heuristics, the simple knowledge of $\mathbf{A}$
(regardless of any permutation of the variable indexes) is enough to find the set
of solutions by using the Gaussian elimination.

A similar strategy is used in the generation of the \emph{Wishart} instances
\cite{hamze2020wishart}. In this case, real valued matrices $\mathbf{W} \in
\mathbb{R}^{ k \times N}$ are chosen such that $\mathbf{W} x^* = 0$ for the
desired binary variable.  Therefore, the corresponding binary cost function
$H(x) = \frac{1}{2} x^T \mathbf{W}^T \mathbf{W} x = \frac{1}{2}\sum_{ij} J_{ij}
x_i x_j$ is guaranteed to have $x^*$ as a solution.  The hardness of the
Wishart instances is tuned by properly choosing the ratio $\alpha = k/N$, with
an easy-hard-easy pattern by increasing $\alpha$.  Interestingly, the study of
the Thouless-Anderson-Palmer (TAP) equations \cite{thouless1977solution}
applied to the Wishart problem shows the existence of a first order phase
transition, making the Wishart instances hard to optimize for local heuristics.
Moreover, unlike many other planting techniques, the Wishart instances are
fully-connected and hard to optimize, even for a small number of variables.
However, the hardness of the Wishart instances also depends on the digit
precision of $\mathbf{W}$, making it less suitable for devices with a limited
precision. Moreover, the number of solutions, which also may affect the
hardness of the Wishart instances, cannot be determined except by an exhaustive
enumeration and cannot be fixed in advance.

In \cite{king2019quantum}, the fact that local heuristics have trouble
optimizing frustrated loops is used to generate random instances with a known
ground state, and hence the name Frustrated Cluster Loops (FCL). The ruggedness
$R$ of an FCL instance is how many loops a single variable is involved in, and
it determines the hardness of the instances. Unfortunately, cluster methods
easily identify these frustrated loops and therefore solve the FCL instances
efficiently \cite{mandra2016strengths, mandra2017pitfalls, mandra2019state}. A
variation of the FCL planting method has been used in
\cite{mandra2018deceptive} to generate hard instances. In this case, FCL
instances are embedded on a Chimera graph with an embedding cost tuned by
$\lambda$. The name Deceptive Cluster Loops (DCL) derives from the fact that
for $\lambda$ small enough, the ground state of DCL instances is mappable to
the ground state of an equivalent 2D model, while for $\lambda$ large to the
ground state of a fully connected model. The hardness is then maximized at the
transition between the two limits. However, the ground state can only be
determined at the extrema of $\lambda$.

Instead of taking the ``shotgun'' approach of generating spin-glass instances
on a native graph until a sufficient number of hard instances are found, the
hardness of the problem itself is treated as a cost function to optimize over
the space of Ising Hamiltonians in \cite{marshall2016practical}, where they
apply a Metropolis update rule on the Hamiltonian couplings to seek out harder
instances. While the resulting instances are significantly harder, this
approach requires an expensive quantification of the problem hardness by
solving a new instance at every step. Furthermore, the ground state energies of
the generated instances are not known in advance, and enforcing compatibility
with a planted solution limits the available space of moves to generate novel
and harder instances.

One of the first examples of the use of a private/public key protocol to
generate hard instances can be found in \cite{merkle1978hiding}. More
precisely, the authors of the paper use the subset sum problem
\cite{sipser1996introduction}, where a subset of a set of numbers $\Omega =
\{\omega_1,\, \omega_2,\ \ldots,\ \omega_N\}$ must be found such that their sum
equals a given value $c$. The problem is known to be NP-Complete, and its
quadratic binary cost function can be expressed as $H(x) =
\sum_{i=1}^n\left(\omega_i x_i - c\right)^2$. However, the subset sum problem
has a pseudo-polynomial time algorithm \cite{papadimitriou2003computational}
which takes $\mathcal{O}(n \omega^*)$ memory, with $\omega^*$ being the largest
value $\omega_i$ in the set. The Merkle-Hellman encryption algorithm works by
using two sets of integers: the ``private'' key $\Omega$, which is a
super-increasing set of numbers, and a ``public'' key $\Omega^\prime$, which is
defined as $\omega^\prime_i = r \omega_i\ \text{mod}\ q$, with $q > \sum_i
\omega_i$ and $r$ being a random integer such that $r$ and $q$ are coprime. The
values $r$ and $q$ are also kept secret, A message $m_i$ of $N$ bits is then
encrypted as $c = \sum_i m_i \omega^\prime_i$.  On the one hand, without the
knowledge of the private key, one must optimize the quadratic binary cost
function $H(x) = \left(\omega^\prime_i x_i - c\right)^2$ in order to find the
encoded message. On the other hand, thanks to the fact that $\Omega$ is a set
of super-increasing numbers and that the inverse of $r$ modulo $q$ can be
efficiently found using the extended Euclidean algorithm, one can easily
recover the binary message without the optimization of $H(x)$.  The fact that
the binary message $m_i$ is recoverable guarantees that the ground
state must be \emph{unique}. Unfortunately, the use of super-increasing
sequence of numbers makes the Merkle-Hellman protocol vulnerable to attacks
\cite{shamir1982polynomial} by just looking at the public key $\Omega^\prime$.
Moreover, the complexity from the computational point of view depends on
$\omega^*$ which sets the minimum digit precision for the set $\Omega$. That
is, exponentially hard integer partition instances require exponentially large
$\omega_i$, limiting the use of this protocol on devices with limited
precision.

Recently, the integer factorization problem has been used to generate hard
binary instances \cite{pirnay2022super, szegedy2022quantum}. More precisely,
given two large prime numbers $a$ and $b$ of $N-$binary digits, the binary cost
function is defined as $H(x,\ y) = \left[\left(\sum_{i=1}^N 2^i
x_i\right)\left(\sum_{i=1}^N 2^i y_i\right) - c\right]^2$, with $c = ab$.
Instances based on factoring become exponentially hard by increasing the number
of binary digits $N$. While no known classical algorithms are known to
efficiently solve $H(x,\ y)$ without actually optimizing the binary cost
function, factoring is vulnerable to quantum attacks via the Shor algorithm
\cite{shor1999polynomial}. Moreover, the optimization of the binary cost
function $H(x,\,y)$ requires an exponential precision, making such instances
unsuitable for devices with limited precision.
 
\section{Background on Linear Codes} \label{sec:linear_codes}

To simplify the reading, for the rest of the paper it is assumed, unless
otherwise specified, that all matrices and linear algebra operations take place
over the finite field $\mathbb{F}_2 \equiv \mathbb{Z}/2\mskip1.5mu\mathbb{Z}$,
\ie module 2 addition and multiplication over bits.  Let us define the
generator matrix $\mathbf{G} \in \mathbb{F}_2^{\,k\times N}$ of a linear code
$C$, with $k \leq N$, and its check matrix $\mathbf{H} \in
\mathbb{F}_2^{\,(N-k)\times n}$ such that $\mathbf{H}\mathbf{G}^T = 0$.  A
message $x \in \mathbb{F}_2^{\,k}$ can be then encoded as a string $y \in
\mathbb{F}_2^{\,N}$, called the codeword, via
\begin{equation}
	y \equiv x\,\mathbf{G}.
\end{equation}
The code rate of $C$ is defined as $R = k/N$, which is the average rate of
useful information (the $k$-bit message) received per bit transmitted with the
code (the $N$-bit codeword).  Codes with smaller rates can incorporate more
redundant bits which makes them resistant to random errors when transmitted
across a noisy channel.  The distance of $C$ is defined as the minimum Hamming
weight (the number of non-zero bits) among all the possible non-zero codewords,
that is:
\begin{align}
	d \equiv \min_{x\in\mathbb{F}_2^{\,k},\ x\not=0} \left\Vert x\,\mathbf{G} \right\Vert,
\end{align}
where $\Vert\cdot\Vert$ indicate the Hamming weight. By the linearity of the
code, the Hamming distance $\Vert y_1 - y_2\Vert$ between any two codewords
$y_1$ and $y_2$ is at least $d$. Thus, a linear code of distance $d$ can detect
up to $d-1$ errors and correct up to $(d-1)/2$ errors.  Indeed, let us define a
vector $\epsilon$ with $t = \left\lvert\epsilon\right\rvert$ ones. The error
would alter the encoded message as:
\begin{align}
	y^\prime = x\,\mathbf{G} + \epsilon
\end{align}
Since $x\,\mathbf{G}\mathbf{H}^T = 0$ by construction, applying the parity
check to $y^\prime$ one gets:
\begin{align}
	y^\prime\,\mathbf{H}^T = \epsilon\, \mathbf{H}^T \equiv z.
\end{align}
where $z$ is called the syndrome of the error.  However, if $t < d$, $\epsilon$
cannot be a codeword of $C$ and therefore $\epsilon\,\mathbf{H}^T \not= 0$,
that is the code $C$ can detect up to $d - 1$ errors. The three parameters $N$,
$k$, and $d$ are the most important characterization of linear codes, so we
conventionally denote them by writing that $C$ is an $[N, k, d]$ linear code.

An error can be instead corrected if, for a given syndrome
$z=\epsilon\,\mathbf{H}^T$, it is possible to unequivocally identify
$\epsilon$.  This is the case when two errors $\epsilon_1$ and $\epsilon_2$
produce two different syndromes. More precisely
\begin{equation}
	\epsilon_1\,\mathbf{H}^T \not= \epsilon_2\,\mathbf{H}^T \Rightarrow
	\left(\epsilon_1 - \epsilon_2\right)\mathbf{H}^T \not= 0,
\end{equation}
that is $\epsilon_1 - \epsilon_2$ is not a codeword of $C$.  However, if
$\epsilon_1$ and $\epsilon_2$ have respectively $t_1$ and $t_2$ errors, their
sum cannot have more than $t_1 + t_2$ errors, implying that two syndromes can
be distinguished only if the number of errors is smaller than $t \leq (d -
1)/2$.

\subsection{McEliece Public/Private Key Protocol}

In 1978, Robert McEliece discovered that linear codes $C$ can be used as
asymmetric encryption schemes \cite{mceliece1978public}. While not getting much
acceptance from the cryptographic community because of its large private/public
key and being vulnerable to attacks using information-set decoding
\cite{bernstein2008attacking}, the McEliece protocol has recently gained more
traction as a ``post-quantum'' cryptographic protocol, as it is immune to the
Shor's algorithm \cite{dinh2011mceliece}. At the time of writing, NIST has
promoted the McEliece protocol to the phase 4 of its standardization
\cite{NIST_PostQ}.

In the McEliece protocol, the private key consists of a linear code $C$ of
distance $d$ (with $\mathbf{G}$ and $\mathbf{H}$ being its generator matrix and
check matrix respectively), a (random) permutation matrix
$\mathbf{P}\in\mathbb{F}_2^{\,N\times N}$, and a (random) non-singular matrix
$\mathbf{S}\in\mathbb{F}_2^{\,k\times k}$. On the other hand, the public key is
defined as a new generator matrix $\mathbf{G}^\prime =
\mathbf{S}\mathbf{G}\mathbf{P}$ with distance $d$. Indeed, $\Vert
x\,\mathbf{G}^\prime\Vert = \Vert x\,\mathbf{S}\mathbf{G}\mathbf{P}\Vert =
\Vert x\,\mathbf{S}\mathbf{G}\Vert$ since the Hamming weight is invariant by
permutation, and
\begin{equation}
	\min_{x\in\mathbb{F}_2^{\,k},\ x\not=0}
	\left\Vert x\,\mathbf{S}\mathbf{G}\right\Vert =
	\min_{y\in\mathbb{F}_2^{\,k},\ y\not=0}
	\left\Vert y\,\mathbf{G}\right\Vert = d,
\end{equation}
where we used the fact that $\mathbf{S}$ is non-singular and $y =
x\,\mathbf{S}$. It is important to notice that it is hard to recover the
original generator matrix $\mathbf{G}$ from $\mathbf{G}^\prime$ without the
knowledge of $\mathbf{P}$ and $\mathbf{S}$. In the McEliece protocol,
$\mathbf{S}$ is used to ``obfuscate'' the original generator matrix
$\mathbf{G}$ by scrambling it. In the next Section, we will use a similar
concept to create hard instances with a known and unique ground state.

The message encoding is then obtained by applying the public key
$\mathbf{G}^\prime$ to an arbitrary message $q\in\mathbb{F}_2^{\,k}$ and
randomly flipping a number $t = (d-1)/2$ of bits, that is:
\begin{equation}\label{eq:mcelieceEncrypt}
	q^\prime \equiv q\,\mathbf{G}^\prime + \epsilon,
\end{equation}
with $\Vert\epsilon\Vert = t$.
If the private key is known, the message $q$ can be recovered by observing that
$\epsilon^\prime = \epsilon\, \mathbf{P}^{-1}$ preserves the number of errors.
Therefore the permuted encoded message $q^\prime \mathbf{P}^{-1} =
q\,\mathbf{S}\mathbf{G} + \epsilon^\prime$ can be ``corrected'' using the
parity check $\mathbf{H}$. Indeed
\begin{equation}
	q^\prime\mathbf{P}^{-1}\mathbf{H}^T = \left(q\,\mathbf{S}\mathbf{G} +
	\epsilon^\prime\right)\mathbf{H}^T = \epsilon^\prime\mathbf{H}^T
\end{equation}
by construction. Because $\Vert\epsilon^\prime\Vert \leq (d-1)/2$, one can
identify $\epsilon^\prime$ and remove it from the permuted message and
therefore recover $q\,\mathbf{S}$. Finally, the encoded message can be
recovered by multiplying $\mathbf{S}^{-1}$ to the right of $q\,\mathbf{S}$.

Without the knowledge of the private key, the message $q^\prime$ can be decoded
by using the maximum likelihood estimation:
\begin{equation}\label{eq:max_likelihood}
	q = \underset{x\in\mathbb{F}_2^{\,k}}{\rm argmin}\left\Vert q^\prime -
	x\,\mathbf{G}^\prime\right\Vert.
\end{equation}
Recalling that Eq.~\eqref{eq:max_likelihood} is invariant by non-singular
transformations $x\leftarrow x\, \mathbf{Q}$ with
$\mathbf{Q}\in\mathbb{F}_2^{\,k\times k}$, it is always possible to reduce
$\mathbf{G}^\prime$ to its normal form $\mathbf{Q}^{-1}\mathbf{G}^\prime =
\left(\mymathbb{1}\  \mathbf{W}\right)$, with $\mymathbb{1}$ being the identity
matrix of size $k\times k$ and $\mathbf{W}\in\mathbb{F}_2^{\,k\times(N-k)}$.

Since the encrypted message $q$ is recoverable, it is guaranteed that one and
only one configuration exists that minimizes Eq.~\eqref{eq:max_likelihood}.
The hardness of the McEliece protocol depends on the size of the codespace and
the distance $d$ of the code $C$. By properly choosing $k$ and $d$, it is
possible to generate instances with a tunable hardness.
 
\section{Generating Hard Instances from Linear Codes}\label{sec:hard_instances}

In this section we will show that the McEliece protocol can be used to generate
hard instances of a disordered Ising Hamiltonian. More precisely, we are
interested in generating \mbox{$p$-local} Ising instances of the form:
\begin{equation}\label{eq:plocal_ising}
	H(\sigma) = \sum_{i=1}^M
	J_i\, \sigma_{i_1}\cdots\, \sigma_{i_p}
\end{equation}
with $p \geq 1$, $1 \leq i_1,\,\ldots,\,i_p\leq k$, $\sigma = \{\pm1\}^k$ and
couplings $J_{i_1 \cdots i_p}\in\mathbb{R}$.
To achieve this goal, it suffices to show that finding a solution of the
maximum likelihood equation in Eq.~\eqref{eq:max_likelihood} is equivalent to
finding the ground state of the \mbox{$p$-local} Ising instance
\eqref{eq:plocal_ising}. This equivalence is well-known in the theory of spin
glasses and has motivated statistical-mechanical studies of various error
correcting codes \cite{sourlasSpinglass89,nishimoriStatistical01}.

Let us define $\mathbf{G}^\prime_i$ the $i$-th column of $\mathbf{G}^\prime$.
Therefore, it is immediate to show that:
\begin{equation}\label{eq:ml_1}
	W(x) = \Vert q^\prime - x\,\mathbf{G}^\prime\Vert =
	\sum_{i=1}^N \Vert q^\prime_i - x\cdot g^\prime_i\Vert,
\end{equation}
where the operations inside $\Vert\cdot\Vert$ are intended modulo $2$, while
the sum is over $\mathbb{R}$. By mapping $\sigma_i = 1-2x_i$, and by recalling
that
\begin{subequations}\label{eq:x_to_sigma}
	\begin{align}
		(x_1 + \cdots + x_k)\!\! \mod 2     & = (1-\sigma_1\cdots\sigma_k)/2, \\
		(x_1 + \cdots + x_k + 1)\!\! \mod 2 & = (1+\sigma_1\cdots\sigma_k)/2,
	\end{align}
\end{subequations}
Eq.~\eqref{eq:ml_1} becomes:
\begin{align}\label{eq:ml_2}
	W(x) & = \sum_{i=1}^N \Vert q^\prime_i - x\cdot \mathbf{G}^\prime_i\Vert =
	\sum_{i=1}^N\left[1 - (-1)^{q^\prime_i}\sigma_{i_1}\cdots\sigma_{i_{p_i}}
	\right]\nonumber                                                                \\
	     & = -\sum_{i=1}^N(-1)^{q^\prime_i}\sigma_{i_1}\cdots\sigma_{i_{p_i}} + N =
	H(\sigma) + N,
\end{align}
with $\{i_1,\,\ldots,\,i_{p_i}\}$ being the indices of the non-zero elements of
$\mathbf{G}^\prime_i$, and $p_i = \lvert g^\prime_i\rvert$ their number. It is
interesting to observe that the indices of the interactions
$\{i_1,\,\ldots,\,i_{p_i}\}$ are fixed after choosing the private key
$\{\mathbf{G},\ \mathbf{H},\ \mathbf{P},\ \mathbf{S}\}$. However, multiple
random instances can be generated by providing different $q^\prime =
q\mathbf{G}^\prime + \epsilon$, which will only change the sign of the
interactions. It is important to emphasize that, if the private key is properly
chosen, every instance $H(\sigma)$ will be hard to optimize, regardless of the
choice of $q^\prime$. This is consistent with the fact that  $p$-local
instances with randomly chosen $J_{i_1 \cdots i_p} = \pm1$ and $p \geq 2$ are
NP-Complete. However, unlike the case when the couplings are randomly chosen,
$p$-local instances generated using the McEliece protocol have a
\emph{guaranteed} unique optimal value, which is known in advance (that is,
$q$).

Since we don't need to apply $\mathbf{S}^{-1}$ to obtain $H(\sigma)$ in
Eq.~\eqref{eq:ml_2}, we can relax the request to have the obfuscation matrix
$\mathbf{S}$ be non-singular. In particular, the number of solutions of the
$H(\sigma)$ will depend on the size of the kernel of $\mathbf{S}$.  Indeed, not
having $\mathbf{S}$ invertible implies that $\mathbf{S}$ has a non-empty kernel
and, therefore, two distinct messages $q_1$ and $q_2$ may have the same
encrypted message, $q^\prime_1$ and $q^\prime_2$. Consequently, the $p$-local
instance $H(\sigma)$ will have as many optimal configurations as the number of
collisions for a given message $q$, that is the number of elements in the
$\mathbf{S}$ kernel.

\subsection{Reduction to the 2-local Ising Model}\label{sec:reduction_2local}

Since the $p$-local Ising model is NP-Complete for $p \geq 2$, any optimization
problem in the NP class can be mapped onto it. However, almost all post-Von
Neumann \cite{mohseni2022ising, sao2019application, traversa2015universal,
adleman1994molecular} technologies can only optimize $p$-local Ising instances
with either $p=2$ or $p=3$.  Unfortunately, because of the random scrambling
induced by $\mathbf{S}$, $\mathbf{G}^\prime$ is a dense matrix with \mbox{$p_i
\leq k$}.  Luckily, $p$-local couplings with $p > 3$ can always be reduced to
$p \leq 3$ by adding auxiliary variables.

For $p>3$, it is immediate to show that:
\begin{multline}\label{eq:plocal_red}
	(-1)^t\sigma_1\cdots\sigma_l\sigma_{l+1}\cdots\sigma_p = \\
	\min_{\omega = \pm1}\left[\sigma_1\cdots\sigma_l\,\omega +
		(-1)^t\,\omega\,\sigma_{l+1}\cdots\sigma_p\right] + 1.
\end{multline}
Therefore, reducing $\mathbf{G}^\prime$ to a $3$-local Ising instance would
require no more than $k^\prime = k + \left(\left\lceil \frac{k}{2} \right\rceil
- 1\right)(N-k)$ spin variables and $n^\prime = k + \left\lceil \frac{k}{2}
\right\rceil(n-k)$ couplings in the worst case scenario.
Reducing a coupling from $p=3$ to $p=2$ requires the addition of another extra
auxiliary spin for each $3$-local coupling. More precisely, since
\begin{multline}
	\underset{\sigma_1,\,\sigma_2,\,\sigma_3=\pm1}{\rm argmin}
    (-1)^t \sigma_1 \sigma_2 \sigma_3 = \\
	\underset{\sigma_1,\,\sigma_2,\,\sigma_3\,\omega=\pm1}{\rm argmin}
	\left[\sigma_1 + \sigma_2 + (-1)^t\left(1 - 2w\right)\right]^2,
\end{multline}
it is possible to reduce $\mathbf{G}^\prime$ to up to $k^{\prime\prime} \equiv
k(N-k+1) = \mathcal{O}(kN)$ spin variables and $N^{\prime\prime} \equiv N + k(n-k) =
\mathcal{O}(kN)$ couplings in the worst case scenario.

\subsection{The Scrambling Protocol}\label{subsection:scrambling_protocol}

In this Section we will show how we can use the ``scrambling'' of the codespace
of $C$ induced by the non-singular matrix $\mathbf{S}$ to generate random hard
$p$-local Ising instances with a known optimal configuration. To this end, it
suffices to show that any $p$-local Ising instance of the form
Eq.~\eqref{eq:plocal_ising} can always be reduced to a maximum likelihood
problem as in Eq.~\eqref{eq:max_likelihood}.

Using the transformation in Eq.~\eqref{eq:x_to_sigma}, the $p$-local Ising
instance in Eq.~\eqref{eq:plocal_ising} can be rewritten as:
\begin{multline}\label{eq:plocal_ising_mlh}
	H(x) = \sum_{i=1}^M {\rm abs}\left(J_i\right)
	\Big(\Big\Vert x_i + \cdots + x_{i_p} + \\
	+ \frac{1 + {\rm sgn}(J_i)}{2} \Big\Vert - 1\Big),
\end{multline}
where ${\rm abs}(\cdot)$ and ${\rm sgn}(\cdot)$ are the absolute and sign
functions respectively, and all operations inside $\lvert\cdot\rvert$ are
\mbox{modulo $2$}. If we define $\mathbf{J}\in\mathbb{F}_2^{\,k \times M}$  the
adjacency matrix of the couplings $J_{i_1 \cdots i_p}$ such that
$\mathbf{J}_{ai} = 1$ if $a \in \{i_1,\,\ldots,\,i_p\}$ and zero otherwise,
Eq.~\eqref{eq:plocal_ising_mlh} becomes
\begin{equation}\label{eq:plocal_ising_mlh2}
	H(x) = \sum_{i=1}^M {\rm abs}\left(J_i\right)
	\Big(\Big\Vert x\cdot\mathbf{J}_i + \frac{1 + {\rm sgn}(J_i)}{2} \Big\Vert - 1\Big),
\end{equation}
where $\mathbf{J}_i$ is the $i$-th column of $\mathbf{J}$ and
$x\in\mathbb{F}_2^{\,k}$.  Observe that Eq.~\eqref{eq:plocal_ising_mlh2} is
equivalent to the maximum likelihood in Eq.~\eqref{eq:max_likelihood} when
${\rm abs}(J_i) = 1$ for all $i$.  More importantly, it is immediate to see
that, for any non-singular transformation, $\mathbf{S}$
an optimal configuration $x^*$ of $H(x)$ is mapped to $x^{*\prime} =
x^*\mathbf{S}^{-1}$, preserving the number of optimal configurations.  We can
relax the condition of having $\mathbf{S}$ being non-singular.  In this case,
the number of optimal configurations is not preserved since
\begin{equation}
	x^* = \left(x^{*\prime} + \xi\right)\mathbf{S}
\end{equation}
for any $\xi$ belonging to the kernel space of $\mathbf{S}$.  Once the
adjacency matrix $\mathbf{J}^\prime = \mathbf{S}\,\mathbf{J}$ is scrambled, one
can map Eq.~\eqref{eq:plocal_ising_mlh2} back to a $p$-local Ising instance using
the methodology presented in Section~\ref{sec:reduction_2local}.

In the next Section we try to shed light on the reason why scrambling the
configuration space $x$ with $\mathbf{S}$ produces instances which are hard to
optimize for local heuristics.

\section{Random Scrambling of Energy Landscapes}\label{sec:clustering}

In order to study the effect of energy scrambling over an Ising Hamiltonian, we
show in this Section how various forms of energy scrambling --- similar to the
one induced by the scrambling matrix $\mathbf{S}$ of the McEliece protocol ---
can transform simple problems into significantly more complex ones by
introducing a version of a phenomenon known in the statistical physics
literature as ``clustering'' \cite{PhysRevLett.94.197205,
10.1145/1132516.1132537}, where the solution space of a constraint optimization
problem splits into a (typically exponential) number of ``clusters''. This
phenomenon is conjectured -- on the basis of empirical evidence collected in
the study of constraint satisfaction problems -- to impair the performance of
stochastic local search algorithms such as WalkSAT
\cite{doi:10.1137/16M1084158} even though the precise connection between
clustering and the theory of algorithmic hardness is still an active area of
research, and not fully understood \cite{doi:10.1073/pnas.2108492118,
PhysRevX.13.021011}.

In order to understand the algorithmic hardness of the system  of Eq.
\eqref{eq:plocal_ising} using the methods of statistical physics, one would like
to study the typical properties of the energy landscape of the ``disordered
Hamiltonian'' $H(\sigma)$ where the disordered couplings $J_{{i_1}\ldots i_p}$
are generated by \eg a fixed choice of the linear code matrix $\mathbf{G}$ (one
for each system size $N$), and the random choices for the permutation matrices
$\mathbf{S},\mathbf{P}$. For this introductory work we limit ourselves to a
simpler scenario that allows us focusing on one component of the McEliece
protocol: the effect of an energy scrambling matrix $\mathbf{S}$. In particular
\begin{enumerate}
  \item We only consider the effect of one source of quenched disorder, \ie a
    random scrambling matrix $\mathbf{S}$. As shown in
    Section~\ref{subsection:scrambling_protocol}, the effect of changing the
    $\mathbf{S}$ permutation matrix in the McEliece protocol (for given
    $\mathbf{G}$ and $\mathbf{P}$) is to shuffle the energies of the states of
    the system.
  \item Instead of considering the effects that $\mathbf{S}$ has on the energy
    landscape of a reference Hamiltonian $H(\sigma)$ generated by fixed
    $\mathbf{G}$ and $\mathbf{P}$, we use as a reference energy landscape a
    simple ``convex'' one that can be easily navigated by steepest descent.
\end{enumerate}
We note that these simplifications, while creating some disconnect with the
McEliece Hamiltonian, have the advantage of making the scrambling protocol of
broader application to more generic Ising spin models. We leave a complete
analysis of the McEliece Hamiltonian for future work.

The main results of this section show that energy scrambling can generate
clustered (empirically ``hard'') landscapes from unclustered (empirically
``easy'') ones. In order to review the concept of clustering, let us assume
that we have some disordered Hamiltonian $H_J$ (dependent on some quenched
disorder $J$) over a system of $N$ Ising spins, so that each $\sigma \in \{\pm
1\}^N$ is assigned an energy $E_\sigma$. For a given energy density $\epsilon =
E/N$ and value of the fractional Hamming distance $x = X/N \in [0,1]$, one
defines the quantity:
\begin{equation}
	\mathcal{N}(x,\epsilon) \equiv \sum_{\sigma,\tau} \delta
    \Big(d_{\sigma,\tau} -xN \Big) \delta(E_\sigma - \epsilon N) 
    \delta(E_\sigma - \epsilon N),
\end{equation}
where the double sum is over all spin configurations $\sigma,\tau \in \{\pm
1\}^N$, $d_{\sigma,\tau}$ is the Hamming distance between $\sigma$ and $\tau$,
$E_{\sigma},E_{\tau}$ are the energies of the configurations, and $\delta(y)$
are Kronecker delta functions (meaning that $\delta(y)=1$ if $y=0$, and
$\delta(y)=0$ otherwise).

For a given realization of disorder, this function computes the number of pairs
of spin configurations \mbox{$\sigma,\tau \in \{\pm 1\}^N$} such that
\begin{enumerate}
	\item $\sigma,\tau$ are exactly $x N$ spin flips away from each other
	\item both $\sigma$ and $\tau$ have the same energy $\epsilon N$
\end{enumerate}
This means that $\mathcal{N}(\epsilon) \equiv \int_0^1 \mathcal{N}(x,\epsilon)
\, \mathrm{d}x$ is a quadratic function of the number of states of given energy
density $\epsilon$. In many spin glass models and constraint satisfaction
problems one expects that the number of states of given energy density should
be of exponential order in $N$, and consequently one should have
\mbox{$\mathcal{N}(x,\epsilon) \sim \exp(N \Phi_{x,\epsilon} + o(N))$} for some
$N$-independent real number $\Phi_{x,\epsilon}$.

The main observation here is that there can exist energy densities $\epsilon$
such that for some values of $x$, \mbox{$\mathcal{N}(x,\epsilon) \rightarrow
0$} as $N \rightarrow \infty$, while for other values $x'$,
$\mathcal{N}(x',\epsilon)$ diverges with $N$. This means that in the
thermodynamic limit, the spin configurations that populate the micro-canonical
shell of energy $\epsilon N$ are not arbitrarily close or far away from each
other, but can exist only at some given distances. Indeed, with increasing $x$
from zero to one, the quantity $\mathcal{N}(x,\epsilon)$ typically has a
diverging region (for $x$ between $0$ and some $x_{1}$), followed by a region
where $\mathcal{N}(x,\epsilon) \rightarrow 0$ (that we call a ``forbidden
region''), followed by a revival of the divergent behavior (from $x_{2} > x_1$
to $1$).

The emerging picture is that these configurations form extensively-separated
``clusters'', where configurations included in the same cluster can be
connected by ``jumping'' along a sequence of configurations (all belonging to
the energy shell), each $\mathcal{O}(1)$ spin flips away from the previous one,
while configurations belonging to different clusters require an extensive
number of spin flips.

In a disordered system the quantity $\mathcal{N}(x,\epsilon)$ will depend on
the disorder realization, so one is led to studying some statistics of
$\mathcal{N}(x,\epsilon)$, the easiest being its disorder-averaged value
\begin{align}
	\langle \mathcal{N}(x,\epsilon) \rangle & =
	\sum_{\sigma,\tau} \delta \Big(d_{\sigma,\tau} -xN \Big) 
    \Big\langle \delta(E_\sigma - \epsilon N) 
    \delta(E_\sigma - \epsilon N) \Big\rangle\nonumber \\
	                                        & = 2^N \binom{N}{xN} 
    \operatorname{Pr}\Big[ E_{\sigma} = \epsilon N, E_{\tau} = \epsilon N \Big],
\end{align}
where we have used the fact that the expectation value of the indicator
function of an event is equal to the probability of that event. Then the
meaning of $\avg{\mathcal{N}(x,\epsilon)} \rightarrow 0$ for $N\rightarrow
\infty$ is that the probability of finding disorder realizations with at least
one pair of states satisfying conditions 1. and 2. above, is vanishing as one
approaches the thermodynamic limit. This is because while the expectation value
of a variable is not necessarily a good indicator of its typical behavior,
since $\mathcal{N}(x,\epsilon)$ is a non-negative random variable one can use
Markov's inequality to show that for any $\alpha >0$
\begin{equation}\label{eq:markov_ineq}
	\operatorname{Pr}\Big[ \mathcal{N}(x,\epsilon) > \alpha \Big] \leq 
    \frac{1}{\alpha} \avg{\mathcal{N}(x,\epsilon)},
\end{equation}
so that when $\avg{\mathcal{N}(x,\epsilon)} \rightarrow 0$ in the large-$N$
limit (\ie in the forbidden regions) then the distribution of
$\mathcal{N}(x,\epsilon)$ is also increasingly concentrated in zero, and
behaves like a Dirac delta in the thermodynamic limit. So the average and
typical behavior coincide (in the thermodynamic limit) at least inside of the
forbidden regions, which is enough for us to establish the existence of
clustering. 

Observe that, in all models we consider, the expectation value $\langle
\mathcal{N}(x,\epsilon) \rangle$ decays exponentially to zero inside of the
forbidden regions, as a function of the system size $N$ in the asymptotic
regime $N \rightarrow \infty$. Therefore, the likelihood of generating an
instance with pairs of states at given distance $x$ and energy $\epsilon$, for
$(x,\epsilon)$ values inside the forbidden region (that is, an ``unclustered''
instance), is also exponentially small.

\subsection{The Hamming-Weight Model}

\begin{figure}[t!]\label{fig:phase_diagram_hwm}
	\centering
	\includegraphics[width=0.9\textwidth]{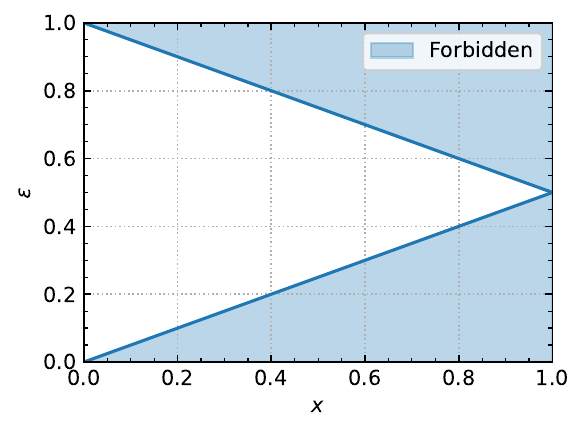}\\
	\includegraphics[width=0.9\textwidth]{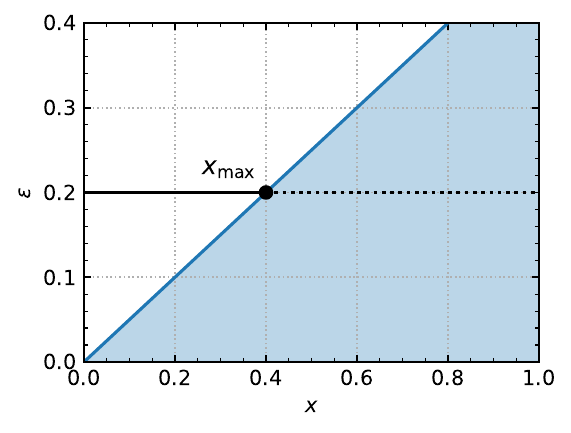}
  \caption{(Top) $(x,\epsilon)$-phase diagram of the HWM. The energy density of
  the HWM is contained in $0 \leq \epsilon \leq 1$, and the fractional Hamming
  distance is $0 \leq x \leq 1$. The regions shaded in blue are given by the
  points $(x,\epsilon)$ where $\mathcal{N}(x,\epsilon)$ is exactly zero at any
  given size. (Bottom) Zoomed-in detail of the $(x,\epsilon)$-phase diagram
  around the energy density $\epsilon = 0.2$. Pairs of configurations with this
  energy density exist at any distance below $x_{\max}=2\epsilon$, and none
  above. This implies that the configurations with this energy density form one
  single cluster of radius $x_{\max}$.}
\end{figure}

As an example, consider the Hamming-Weight Model (HWM) defined by the
Hamiltonian~%
\footnote{by changing into the binary representation $X = (x_1,\ldots,x_N) \in
\{0,1\}^N$ of the spin configuration $\sigma$ through the defining $x_i =
(1-\sigma_i)/2$, then $H(\sigma)$ is the Hamming weight of the associated
binary string $X$.}:
\begin{equation}
	H(\sigma) = -\frac{1}{2}\sum_{i=1}^N (\sigma_i-1).
\end{equation}
This is a non-disordered Hamiltonian and $\mathcal{N}(x,\epsilon)$
can be computed directly. Note that the triangle inequality for the Hamming
distance implies that unless we have that $x/2 \leq \min(\epsilon,1-\epsilon)$,
then $\mathcal{N}(x,\epsilon)=0$. Assuming this inequality to hold, then
\begin{align}
	\mathcal{N}(x,\epsilon) & = \binom{N}{\epsilon N} \binom{\epsilon N}{\frac{1}{2}xN} 
    \binom{(1-\epsilon)N}{\frac{1}{2}xN} \nonumber \\
	                        & =\exp\Big\{N \Big[ S(\epsilon) + 
    \epsilon S\Big(\frac{x}{2\epsilon}\Big) + \nonumber \\
	                        & \hspace{1cm} +
    (1-\epsilon)S\Big(\frac{x}{2(1-\epsilon)}\Big) \Big] + o(N)\Big\},
\end{align}
where we have used Stirling's approximation to show that for $0 \leq p \leq 1$
one has $\binom{N}{p N} = \exp(N S(p) + o(N))$ where $S(p) \equiv -p \ln(p)
-(1-p)\ln(1-p)$ is the binary Shannon entropy.

Note that in this case the $\mathcal{O}(N)$ prefactor $\Phi_{x,\epsilon}$ at
the exponent is
\begin{align}
	\Phi_{x,\epsilon} & = \lim_{N\rightarrow\infty} 
    \frac{1}{N} \ln\mathcal{N}(x,\epsilon) \nonumber \\
	                  & = S(\epsilon) + \epsilon S\Big(\frac{x}{2\epsilon}\Big)
    + (1-\epsilon)S\Big(\frac{x}{2(1-\epsilon)}\Big),
\end{align}
which is never negative since $0 \leq S(p) \leq \ln 2$. Notice that even though
the existence of values $x,\,\epsilon$ where $\mathcal{N}(x,\,\epsilon)=0$ does
mean that there are forbidden regions in the \mbox{$(x,\,\epsilon)$-phase}
diagram of this model, the presence of these regions does not entail
clustering. This is because for any given value of $\epsilon$, the forbidden
region starts at some $x_{\max}>0$ and then extends all the way to $x=1$. This
implies that the configurations with energy $\epsilon N$ cannot exist at a
separation larger than $x_{\max}$ but appear at all distances below that, \ie
they form one single cluster of radius $x_{\max}$. Therefore the HWM does not
exhibit clustering (see Fig.~\ref{fig:phase_diagram_hwm}).

\subsection{The Randomly-Permuted Hamming-Weight Model}

The Randomly-Permuted Hamming-Weight Model (RPHWM) is a disordered system
obtained by taking the HWM and giving a random permutation to the
configurations, while keeping their original energies fixed. Formally, for a
fixed number of spins $N$, one samples a random element $\pi \in S_{2^N}$ of the
permutation group over $2^N$ objects, uniformly at random. Then defines the
energy $E_\sigma$ of a given spin configuration $\sigma \in \{\pm 1\}^N$ as
$E_{\sigma} \equiv \lVert \pi(\sigma) \lVert$ where $\lVert \cdot \lVert$ is
the Hamming weight norm. This model was introduced in \cite{1010.0009}.

\begin{figure}\label{fig:phase_diagram_rphwm}
	\centering
	\includegraphics[width=0.90\textwidth]{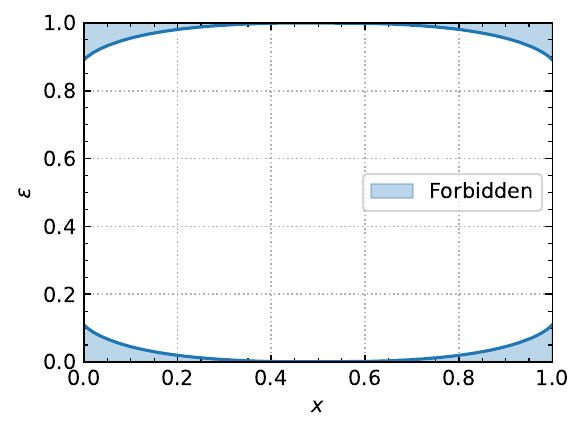}\\
	\includegraphics[width=0.92\textwidth]{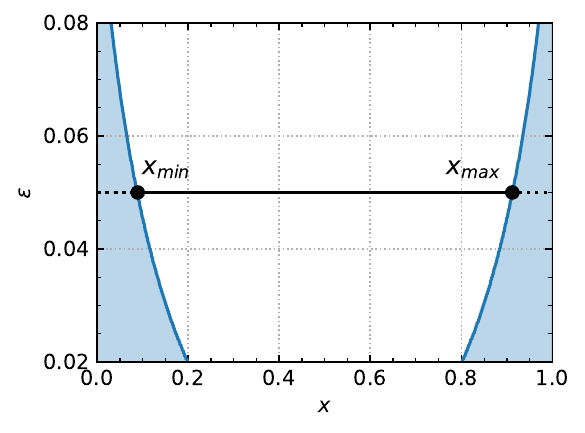}
  \caption{(Top) $(x,\epsilon)$-phase diagram of the RPHWM. The regions shaded
  in blue are given by the points $(x,\epsilon)$ where
  $\mathcal{N}(x,\epsilon)$ goes to zero in the thermodynamic limit. (Bottom)
  Zoomed-in detail of the $(x,\epsilon)$-phase diagram around the energy
  density $\epsilon = 0.05$. Pairs of configurations with this energy density
  cannot exist at a distance smaller than $x_{\min}$ or larger than $x_{\max}$.
  Thus, $x_{\min}$ can be interpreted as the smallest separation between the clusters,
  and each cluster has zero diameter (both in fractional Hamming distance).}
\end{figure}

Sampling the disorder ensemble of the RPHWM can be recast as a process of
extraction from an urn without replacement, where the $2^N$ spin configurations
are given some canonical (\eg lexicographic) ordering
$S^{(1)},\ldots,S^{(2^N)}$, and the urn contains all the possible values of the
energies of the HWM with their multiplicities. These energies are randomly
assigned to the configurations by extracting them sequentially from the urn,
the first energy going to the configuration $S^{(1)}$, the second to $S^{(2)}$
and so on. Following this equivalent reformulation, then one can easily compute
\begin{align}
	\langle \mathcal{N}(x,\epsilon) \rangle & = 2^{N} \binom{N}{xN}
    \frac{1}{2^{N}}\binom{N}{\epsilon N}\frac{1}{2^{N}-1}
    \Big[\binom{N}{\epsilon N}-1\Big]\nonumber \\
	                                        & \hspace{-0.8cm} =
    \frac{1}{2^N-1} \binom{N}{x N}\binom{N}{\epsilon N}^2 - 
    \frac{1}{2^N-1} \binom{N}{x N}\binom{N}{\epsilon N}\nonumber \\
	                                        & \hspace{-0.8cm} \sim
    \exp\Big[ N \Big(-\ln(2) +S(x) +2S(\epsilon)\Big) + o(N)\Big].
\end{align}
Now we have that
\begin{equation}
	\Phi_{x,\epsilon} = -\ln(2) +S(x) +2S(\epsilon)
\end{equation}
is negative iff
\begin{equation}
	S(\epsilon) < \frac{\ln(2)-S(x)}{2}.
\end{equation}
This inequality can be solved numerically to obtain a plot of the forbidden
regions in the $(x,\epsilon)$-phase diagram of the RPHWM (see
Fig.~\ref{fig:phase_diagram_rphwm}). Setting $S(x)=0$ we obtain that forbidden
regions start appearing at energy densities $\epsilon$ that satisfy
$S(\epsilon) < \ln(2)/2$ which for low-energy states happens around $\epsilon
\approx 0.11$.

Notice that to properly establish clustering, one would normally need to
perform second-moment calculations in order to rule out the possibility that
the forbidden regions at a given energy density $\epsilon$ span the
\emph{entire} interval $0 \leq x \leq 1$ (first-moment calculations can be used
to prove the absence of states for given $(x,\epsilon)$ but not their
presence). Such a case would not indicate clustering, but simply the absence of
states at that given $\epsilon$ for the model under study. In our case
second-moment calculations are unnecessary since we know that scrambling --
being a permutation of the states -- leaves unchanged the number of pairs of
states one can have at a given energy density $\mathcal{N}(\epsilon) =
\int_0^{1} \mathcal{N}(x,\epsilon) \, \mathrm{d}x$, which we can compute in the
HWM and is always positive for $\epsilon \in [0,1]$. Thus, the absence of
states at energy density $\epsilon$ is impossible, and the existence of
forbidden regions implies clustering.

Notice, moreover, that the clusters will typically include only a
sub-exponential number of states (their diameter being zero in fractional
Hamming distance \footnote{this is observed also in \eg the Random Energy
Model, where $\Phi_{x,\epsilon} = \ln(2) + S(x) -2\epsilon^2$}) and therefore
most of the spin variables will be \emph{frozen}: starting from a state in a
cluster, flipping any such variable will necessarily move you out of the
cluster. The presence of an extensive number of frozen variables in the set of
solutions has been proposed as the reason behind the hardness of constraint
satisfaction problems over random structures
\cite{PhysRevE.76.031131,Semerjian2008,10.1145/1132516.1132537}. Here we do not
attempt to develop a formal connection between frozen variables and the
algorithmic hardness of our model, but leave such considerations for future
works.

\subsection{The Linearly-Scrambled Hamming-Weight Model}
Let us define how one generates the energies of the states in a disorder
realization of the Linearly-Scrambled Hamming-Weight Model (LSHWM). This is
more easily described by passing to the binary representation of the spin
configurations, \ie by defining the variables $x_i \equiv
(1-\sigma_i)/2$ so that each spin configuration $(\sigma_1,\ldots,\sigma_N) \in
\{\pm 1 \}^N$ is mapped to a binary string $(x_1,\ldots,\,x_N) \in
\mathbb{F}_2^{\,N}$ (see Eq.~\eqref{eq:x_to_sigma}).  In order to generate a
disorder realization of the LSHWM for a system of $N$ spins, one samples $N^2$
numbers $a_{ij} \in \{0,1\}$ with $1 \leq i,j \leq N$ according to the
symmetric probability distribution $\operatorname{Pr}[a =0]=\operatorname{Pr}[a
=1]=1/2$, and defines the ``scrambling matrix'' $\mathbf{S}\in
\mathbb{F}_2^{\,N\times N}$ by using the $\{a_{ij}\}$ as its entries:
$\mathbf{S}_{ij} = a_{ij}$.  Then every string $X = (x_1,\ldots, x_N)$, taken
as a column vector in $\mathbb{F}^N_2$, is mapped to some new string $X^\prime
\in \mathbb{F}_2^{\,N}$ according to the rule
\begin{equation}
	\begin{bmatrix}
		x'_1   \\
		\vdots \\
		x'_N
	\end{bmatrix} = \mathbf{S} \begin{bmatrix}
		x_1    \\
		\vdots \\
		x_N
	\end{bmatrix},
\end{equation}
(where matrix-vector multiplication is defined in $\mathbb{F}_2$) and then we
define the energy $E_{\sigma}$ (that is, the energy of the spin configuration
$\sigma$ represented by the binary string $X$) to be the Hamming weight of the
vector $X^\prime$.

In the Appendix we compute, for $x,\,\epsilon > 0$, the probability
\begin{equation}
	\operatorname{Pr}\Big[E_{\sigma}=\epsilon N, E_{\tau}=\epsilon N\Big] \equiv
    P_{\epsilon} = 2^{-2N} \binom{N}{\epsilon N}^{\!2}
\end{equation}
for two configurations $\sigma,\, \tau \in \{\pm 1\}^N$, neither of which is
the ``all-spins-up'' configuration (represented by the zero vector in
$\mathbb{F}_2^N$, for which a positive energy density is forbidden by
linearity). Then it becomes easy to show that
\begin{align}
	\avg{\mathcal{N}(x,\epsilon)} & = \sum_{\sigma,\tau} 
    \delta\Big(d_{\sigma,\tau}-xN\Big) \operatorname{Pr}\Big[E_{\sigma} = 
    \epsilon N, E_{\tau}=\epsilon N\Big]\nonumber \\
	                              & \hspace{-0.8cm} =
    \sum_{\sigma \neq 0} \sum_{\tau \neq 0} \delta\Big(d_{\sigma,\tau}-xN\Big)
    P_{\epsilon}\nonumber \\
	                              & \hspace{-0.8cm} = 
    (2^N-1)\Big[\binom{N}{xN}-1\Big] P_{\epsilon}\nonumber \\
	                              & \hspace{-0.8cm} \sim 2^N
    \binom{N}{xN} P_{\epsilon}\nonumber \\
	                              & \hspace{-0.8cm} =
    \exp\Big[N \Big(- \ln(2) + 2S(\epsilon) + S(x) \Big) + o(N) \Big]
\end{align}
therefore
\begin{equation}
	\Phi_{x,\epsilon} = 2S(\epsilon)+ S(x) - \ln(2),
\end{equation}
which is negative if and only if
\begin{equation}
	S(\epsilon) < \frac{\ln(2)- S(x)}{2},
\end{equation}
exactly as in the RPHWM. Therefore their clustering phase diagrams are the same.

Note that unlike the previous case, the particular ensemble of scrambling
matrices used here includes non-invertible matrices and, as a consequence, the
uniqueness of the ground state of the model is not guaranteed. However,
provided the typical rank of the scrambling matrix scales like
$\mathcal{O}(N)$, then the fraction of ground states
$2^{N-\operatorname{rank}(\mathbf{S})}/2^N$ in the system is exponentially
decaying with $N$, so these cannot be easily found just by \eg a polynomial
number of random guesses. We refer to Appendix \ref{appendix:kernel} where we
summarize the results the literature, which show that in the limit of large-$N$
the typical nullity of our scrambling matrices converges to a finite value.
Thus, by the rank-nullity theorem, their typical rank must be $\mathcal{O}(N)$. 

\section{Simulations of Hard Scrambled Instances}\label{sec:numerics}

With an understanding of why local search heuristics are ineffective for a
scrambled Hamiltonian, we will now aim to measure the hardness of scrambled
instances generated with a cryptographically realistic family of public/private
key pairs for the McEliece protocol. While the RPHWM may appear statistically
hard, a more clever optimizer can use the knowledge that the underlying
Hamiltonian is the HWM to directly calculate $\mathbf{S}$ from the terms of the
scrambled Hamiltonian with some simple linear algebra and completely side-step
the need to directly optimize it .  Thus, it is necessary to consider a family
of \emph{cryptographically hard} linear error correcting codes which (loosely
defined for our purposes) posses a generator matrix $\mathbf{G}$ that cannot be
easily recovered from its scrambled generator matrix $\mathbf{G'}$ when used in
the McEliece protocol, even with knowledge of how $\mathbf{G}$ could be
generated.  In other words, algorithmic backdoors to optimizing a scrambled
Hamiltonian are prevented by the very cryptanalysis that quantifies how
difficult it is to break the McEliece protocol (\ie solving for the message
$q$ in Eq.~\eqref{eq:max_likelihood}) using the best known attacks.

We will first briefly review the overall approach most often applied for
breaking the McEliece protocol, focusing on Stern's algorithm and its success
probability. We will then present numerical benchmarks comparing the scaling of
a local search algorithm against Stern's algorithm. We empirically show that
local search on cryptographically hard McEliece instances exhibit an
exponential scaling $O(2^{k})$ for the time-to-solution, no better than
exhaustive enumeration and consistent with the presence of clustering.  Since
$N$ and $k$ are related by the code rate $R=k/N$, local search has an
equivalent scaling of $O(2^{RN})$, which is asymptotically $O(2^N)$ in the
worst case that $R\to 1$ as $N\to\infty$ (as is the case in this section).  On
the other hand, the exponential scaling of Stern's algorithm is $O(2^{c N})$,
where $c=0.05564$ is the theoretical worst-case exponential constant (for any
underlying code rate $R$) and we obtain a numerical scaling consistent with
this exponent.  We establish here a concrete benchmarking parameter regime
using cryptographically hard Ising instances with a demonstrably ample gap to
close between algorithms and Ising devices relying on local search and the
cryptanalytic state-of-the-art.

\subsection{Information Set Decoding and the Stern Algorithm}

An \emph{information set} of an $[N, k, t]$ code is a subset $I\subset
\{1,\ldots,N\}$ of size $k$ such that $G_{|I}$, the square sub-matrix formed by
the columns of $G$ indexed by $I$, is invertible.  That is, if no errors exist
on the bits in the information set, then the message is successfully decoded
using only $G_{|I}$.  A \emph{redundancy set} is a subset $R\subset
\{1,\ldots,N\}$ of size $N-k$ such that $H_{|R}$, the square sub-matrix of
columns of $H$ indexed by $R$, is invertible.  Said similarly, if all of the
errors occur on a redundancy set, then they can be uniquely identified.
Clearly, for any two disjoint sets $I$ and $R$ such that $I\cup R = \{1,\ldots,
N\}$, $I$ is an information set if and only if $R$ is a redundancy set. 

The algorithmic approach of finding an error-free (or nearly error-free)
information set to decode a message is called \emph{information set decoding}
(ISD).  Such algorithms constitute a general attack on cryptosystems based on
linear codes, as they make no assumptions concerning the structure of the
underlying code.  After the generator matrix $\mathbf{G}'$ is mapped to a
$p$-local Hamiltonian via Eq.~\eqref{eq:ml_2}, an ISD algorithm can be seen as
an Ising optimization algorithm that is based on searching for a partition of
the summed Ising terms such that all unsatisfied couplings (or as many
unsatisfied couplings as possible) in  Eq.~\eqref{eq:ml_2} belong to the
redundancy set.  The most na\"ive but illustrative ISD algorithm is to simply
select $k$ out of $N$ bits at random from the encoded message and check if they
are an error-free information set by attempting Gaussian elimination.  Suppose
that we are trying to decode a message with $t$ errors (typically
$t=\lfloor(d-1)/2\rfloor$, the maximum error-correcting capacity of the code).
The probability that this random selection is successful is
\begin{equation}
    P_{\mathrm{succ,ISD}} =
     \frac{{{N-t}\choose{k}}}{{{N}\choose{k}}}
     \,.
\end{equation}
Since the distance of a code can grow with its size, the maximum correctable
distance $t$ does not remain fixed (potentially $t=O(N)$), so the worst-case
cost increases as $O(2^{cN})$ where $c$ is a constant that generally depends on
the code rate $R=k/N$ as $N\to\infty$ for the particular family of codes being
decoded.  However, $c$ is maximized over all code rates $0<R<1$ to yield the
worst-case exponential scaling with $N$ for a specific algorithm using any
underlying code for the McEliece protocol, regardless of its rate $R$ or its
error correcting distance $t$ \cite{beckerDecoding12}.

Stern's algorithm \cite{sternMethod89,canteautNew98} is a refinement of ISD
that relaxes the requirement that all of the error bits bits are found in the
redundancy set; instead, only $t-2p$ of the errors are required, where $p\geq
1$ is an integer parameter of the algorithm (not to be confused with the
indefinite $p$ of a $p$-local Hamiltonian). The remaining $2p$ errors are
searched for by splitting the information set into two random partitions ($I_1$
and $I_2$) and then computing linear combinations of columns consisting of
exactly $p$ columns from $I_1$ and $p$ columns from $I_2$.  An iteration of
Stern's algorithm is thus successful if exactly $2p$ error bits are assigned to
the information set, and then exactly $p$ error bits are assigned into each of
the two partitions.  Thus, the success probability of one iteration of Stern's
algorithm can be calculated to be
\begin{equation}
    P_{\mathrm{Succ,Stern}} = \frac{{t\choose 2p}{{N-t}\choose{k-2p}}}{{N\choose k}} 
                              \frac{{k/2\choose p}^2}{{k\choose 2p}}
    \,.
\end{equation}
Stern's algorithm is generally practical with either $p=1$ or $2$.  A detailed
description of Stern's algorithm and other ISD algorithms is given in
Appendix~\ref{appendix:isd}.  While there have been more sophisticated
combinatorial refinements of ISD in recent literature, Stern's algorithm is
almost as simple to implement as plain ISD, and all further improvements
proposed to date have only resulted in marginally improved worst-case scaling
from $O(2^{0.05564\,N})$ for Stern's algorithm to just $O(2^{0.04934\,N})$ at
the cost of additional--but substantial--complexity to implement
\cite{beckerDecoding12}. 

It should be emphasized that this scaling behavior relies on a choice of code
family that does not have enough structure to infer from the public key, but at
the same time should have an efficient decoding algorithm up to $t$ errors
using the private key. The most successful such family is the Goppa codes,
whose codewords are based on randomly selected elements of a finite field.
This is the very same family originally proposed by McEliece
\cite{mceliece1978public} and has resisted the discovery of a breaking attack
to this day \cite{bernstein2008attacking}.

\subsection{Benchmarking Parallel Tempering Monte Carlo on McEliece Instances}

Having established the asymptotic scaling of the best known cryptographic
attacks on secure implementations of the McEliece protocol, we will now examine
how parallel tempering (PT) Monte Carlo \cite{swendsen1986replica,
marinari1992simulated, hukushima1996exchange}, one of the most effective
heuristics with local-search moves for Ising spin glasses, performs at finding
the ground state of the corresponding scrambled Hamiltonian (\ie McEliece
instances). As summarized above, we will show that the scaling of PT is
substantially separated from the theoretical scaling of an ISD attack on the
cryptosystem and interpret this separation as evidence of clustering.

\emph{Computational Implementations.} 
We randomly generated McElience instances, defined by the scrambled
$\mathbf{G}'$ matrix and encoded $q'$ vector, as described in
Appendix~\ref{appendix:generator}. Every instance consists of both a random
Goppa code and a random binary message of length $k$.  We developed our
implementations of Stern's algorithm, which uses the parity check matrix
$\mathbf{H}$ and syndrome $z$ of the McEliece instance as its input, as well as
PT Monte Carlo for sampling a binomial ($J=\pm 1$) $p$-local Ising Hamiltonian,
using $\mathbf{G'}$ and $q'$ as input directly.  Each replica of PT Monte Carlo
directly samples the $p$-local Hamiltonian with Metropolis Monte Carlo moves,
and not the reduction of the $p$-local Hamiltonian to a 2-local Hamiltonian. We
do this to ensure we are close to the ``best case'' scenario of resource
requirements for a local algorithm. Details of the algorithm are given in
Appendix~\ref{appendix:isd}.  Our implementations are written in C++ and
target x86 CPU architectures. Since iterations of Stern's algorithm are easy
to parallelize, we also included support for MPI parallel workers. We
performed all tests on identical Sandy Bridge Xeon E5-2670 processors and
compiled our code using the Intel 2020 compiler with \texttt{-O3 -xAVX}
optimization flags.
     
The original algorithm described by Stern includes an additional step that
peeks at the sum of the selected columns on a small number of random rows and
checks that it is zero before summing all $N-k$ rows.  Theoretically, this
significantly reduces the number of operations per iteration for a relatively
small reduction in the success probability.  However, we have omitted this step
from our implementation as it adds a substantial amount of indirection within a
hot loop and reduces the opportunity for vectorization when compiling to
optimized CPU instructions, which are especially critical when vectors and
matrices over $\mathbb{F}_2$ are bit-packed in memory.  The performance of the
original Stern algorithm was clearly not optimal for the sub-cryptographically
secure parameter ranges we are interested in. 

Our benchmark metric is the time-to-solution (TTS) at the 99\% confidence level, \ie
\begin{equation}
    \mathrm{TTS} = \tau \frac{\log(1-0.99)}{\log(1-P_{\mathrm{succ}})}\,,
\end{equation}
where $\tau$ is the ``cost'' of each iteration, which can be measured in either
the number of ``elementary bit operations'' or the CPU time per iteration.  Our
theoretical TTS of Stern's algorithm simply takes $\tau$ to be the sum of the
worst-case complexities of its two major stages: $(N-k)^3$ operations for
inversion by Gaussian elimination plus $(N-k)
\left((k/2)\mathrm{C}{p}\right)^2$ operations for the combinatorial search for
$2p$ error bits, yielding $\log_2 \mathrm{TTS}$ in bits of security. The formal
analysis of the computational cost of cryptographic attacks is an ongoing area
of development \cite{bernsteinCryptAttackTester23}.  To measure the TTS of our
implementation of Stern's algorithm, we take $\tau$ to be the average measured
CPU time per iteration and evaluate $P_{\mathrm{succ}}$ across all iterations
on a single instance and estimate the error by the standard error of the mean
TTS over all instances.  The TTS of PT is evaluated by running PT until the
ground state energy is found for 10 repetitions per instance, or up to $10^7$
sweeps, ranking the total runtime required to find the solution of every run
and finding the 99\% quantile across all repetitions and instances; the error
estimate for the TTS is obtained by repeating this over bootstrap samples from
the runtimes being ranked.

\begin{figure}[h]
    \centering
    \includegraphics[width=0.98\linewidth]{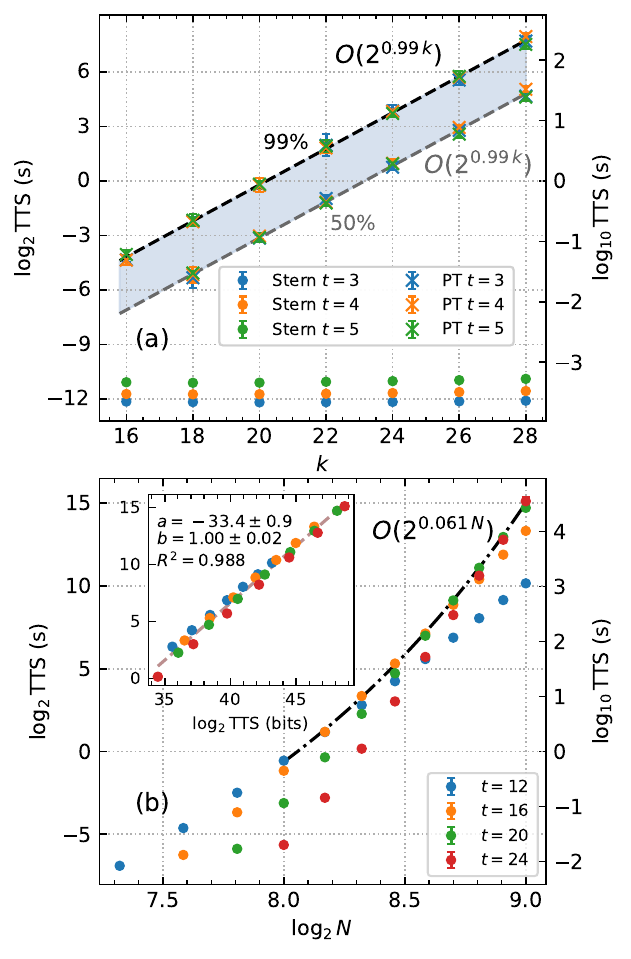}
    \caption{Benchmark results of PT and Stern's algorithm for solving Classic
             McEliece instances constructed from scrambled Goppa codes.  (a)
             Comparison of PT and Stern's algorithms at small error distances
             $t\in\{3,4,5\}$. 
             The 99\% and 50\% quantiles of PT runtimes across all instance
             repetitions are both shown, with the area between the quantiles
             shaded. Both quantiles exhibit an exponential scaling of almost
             exactly $O(2^k)$, independently of code distance. The TTS (99\%)
             of Stern's algorithm rapidly outperforms PT
             (b) Performance of Stern's
             algorithm at $N$ sufficiently large to observe scaling behavior,
             with error distances $t\in\{12,16,20,24\}$.  The direct
             correlation between the measured and theoretical log-TTS of
             Stern's algorithm with a linear fit $(\text{Measured
             log-TTS})=a+b(\text{Theoretical log-TTS})$ is also verified
             (inset) The maximally hard code distances trace an exponential
             envelope of $O(2^{0.061 N})$.  Most error estimates for Stern's
             algorithm are smaller than the plot marker size.  All standard
             errors not specified are within $\pm 2$ at the last significant
             digit.  The finite fields of the Goppa codes were selected to
             avoid jumps in runtime ($\mathbb{F}_2^8$ in (a) and
             $\mathbb{F}_2^9$ in (b)).\label{fig:stern}}
\end{figure}

The timing and TTS results are summarized in Fig.~\ref{fig:stern}.  For
our first set of benchmarks, we examined a parameter regime where the TTS of PT
was measurable with tractable computational effort and its scaling could be
extracted.  We selected a finite field dimension of $m=8$ and code sizes $N$
such that $16\leq k\leq 28$ across modest error-correcting distances
$t\in\{3,4,5\}$ and generated 64 instances for each parameter combination.

In Fig.~\ref{fig:stern} (a), we observe a clear $O(2^{k})$ scaling for PT that
is virtually independent of the code distance, no better than the complexity of
exhaustive enumeration. Even for as few as $k=16$ spins, Stern's algorithm is
more competitive by a factor of about 100, though we performed $10^6$
iterations of Stern's algorithm for each instance to accurately measure the
TTS.  Furthermore, adjusting the code rate would only modulate the ratio of
terms to variables in the Hamiltonian, which only has a polynomial effect
through the time per iteration.  Therefore, the TTS for PT is scaling as the
worst-case scenario, that is proportionally to $2^k$.  While the complexity of
PT is measured as function of $k$, this should be translated to a worst-case
asymptotic as a function of the code size $N$ for any possible that could be
used with the McEliece protocol, regardless of its distance $t$ or its code
rate $R$. Since $R=k/N$ can be arbitrarily close to 1 in the worst case, this
means the worst-case asymptotic of PT is $O(2^N)$, fully exponential in the
code size.  While selection of $m$ was slightly larger than necessary to ensure
that a broad parameter regime was accessible for scaling analysis, we
ultimately limited the error correcting distance to $t\in\{3,4,5\}$ as neither
algorithm behaves characteristically different in this range of $k$.

Fig.~\ref{fig:stern} (b) focuses on a more challenging parameter range for
Stern's algorithm using Goppa codes with $m=9$, code sizes $160\leq N \leq
512$, and a selection of $t\in\{12,16,20,24\}$, which correspond to $40\leq k
\leq 404$ and are computationally inaccessible to PT. For this parameter range,
we generated 32 instances for each parameter combination for the Goppa code.
We performed between $1.6\times 10^6$ and $51.2\times 10^6$ iterations of
Stern's algorithm (distributed over 16 to 64 single-core MPI processes) for
each instance as necessary to reduce the error estimates for increasing code
sizes. We can see the worst-case exponential growth of the TTS by finding the
hardest values of $t$ for each code size.  The measured exponential envelope of
$O(2^{0.061 N})$ scales slightly faster than the theoretical $O(2^{0.05564 N})$
worst-case complexity of Stern's algorithm, though not unexpectedly so since
the TTS is almost, but not fully, in the asymptotic regime.  Not only does
Stern's algorithm extend to a larger order of magnitude for $k$ within a
similar TTS as PT, but we have verified that its exponential scaling is far
more tame.
    
We also observe a strong correlation between the predicted and measured TTS,
verifying that our implementation correctly follows the success probability of
Stern's algorithm. This makes it possible to make a benchmark-driven prediction
of the computational resources required for an implementation of Stern's
algorithm to succeed with a 99\% probability, using a reasonable accounting
method of bit operations.  For example, a McEliece cryptosystem with a
theoretical log-TTS of $64$ bits will require about $2\times10^4$ CPU-days to
be broken with our implementation.  This is not by any means a notable
cryptographic benchmark.  To emphasize the cryptographic hardness of the
problem for much larger $N$, the smallest parameter set proposed for the
Classic McEliece KEM is defined with $N=3488$, $t=64$, and a Goppa code finite
field dimension of $m=12$, which has a theoretical log-TTS of 159.7 bits. Even if
parallelized over an exascale supercomputer cluster such as Frontier, which has
a theoretical peak performance of 1,679.82 petaflop/s, over $10^{22}$ years
would still be required for Stern's algorithm to succeed.  Modest refinements
of Stern's algorithm reaching $O(2^{0.05 N})$ scaling would not prevent an
unrealistically large TTS. In any case, cryptographically hard McEliece
instances establish a clear and rigorous CPU-timed upper bound for the
worst-case performance of local search through Markov chain Monte Carlo on any
class of scrambled problems.

Finally, recall our remark that the mapping of $\mathbf{G}'$ to a scrambled
Ising Hamiltonian yields an algorithmic picture where ISD is partitioning the
Ising terms until the errors (unsatisfied terms in the ground state) can be
singled out. Stern's algorithm thus works in a picture where we search for
$n-k$ terms of the Hamiltonian that contain most of the unsatisfied terms,
perform a gauge transformation of the Hamiltonian (via Gaussian elimination) to
extract a guess state, then attempt to check if it is $2p$ \emph{cluster moves}
(whose size scale with $k$) away from the ground state. The $O(2^k)$ scaling of
PT that we have found is therefore consistent with a complete breakdown of
local search due to clustering in the energy landscape of the Ising that
separates the ground state from a typical excited state by an exponential
number of moves. While Stern's algorithm still scales exponentially, its
complexity has a combinatorial origin and is not impeded by clustering.  

Currently, the best quantum algorithm attack proposed to attack the McEliece
cryptosystem is based on using Grover's algorithm to speed up naive ISD by the
usual Grover speedup \cite{bernsteinGrover10}. However, without scalable
universal fault-tolerant quantum computation, we do not expect any meaningful
benchmarks of this approach to be possible in the near future using McEliece
cryptographic instances. With present non-fault tolerant quantum devices, it is
likely that only QAOA and quantum annealing would be the most suitable quantum
optimization algorithms for such benchmarks. However, variational quantum
algorithms such as QAOA are well-known to end up stuck optimizing in ``barren
plateaus'' when implemented in noisy hardware \cite{wangNoiseinduced21}. Since
Ising problems with clustering are already very unstructured, we expect they
will also be problematic for this class of quantum algorithms. Finally,
currently available quantum annealing hardware is limited to Ising Hamiltonians
with 2-local couplings, which requires the use of additional qubits not only
for the reduction from a $p$-local problem, but also for the minor-embedding of
dense Hamiltonians. Nevertheless, investigating how to overcome these
obstacles and develop a careful scaling analysis of these quantum algorithms
is a suitable subject for future work. 

\section{Conclusion}

In this work we have presented two main results. First, we have proposed a
novel way of randomly generating instances of disordered Ising Hamiltonians
with unique planted ground states by casting the public key of the McEliece
cryptographic protocol as an interacting system of Ising spins. The algorithmic
hardness of finding the ground state of the resulting Hamiltonians is
equivalent to breaking the associated McEliece cryptosystem by only knowing its
public key, while the private key allows the party who knows it to recover the
planted ground state.

Secondly, in order to build some physical intuition as to the reason why our
Hamiltonian model is hard to optimize beyond computational hardness assumptions
of post-quantum cryptography, we have studied the effect of random
energy-scrambling on easy energy landscapes, following the notion that in the
McEliece cryptosystem the public and the private keys are connected through
random permutations. We have shown in simplified settings that even if the
energy landscape of the Hamiltonian induced by a private key has an
easy-to-find ground state, then the energy-scrambling can introduce the
phenomenon known as ``clustering'' into the energy landscape of the Hamiltonian
induced by the public key. The clustering phenomenon is known for being an
obstacle to local stochastic search algorithms.

In order to corroborate this picture, we studied numerically the performance of
Parallel Tempering on Ising instances obtained from the McEliece protocol
through our map -- as an example of the stochastic local search approach -- and
the dedicated Stern algorithm crypto-attack on the original McEliece instances
on the other.  Indeed we observe much faster exponential TTS in the first case
than for the second, confirming that in the absence of information about the
structure of the underlying cryptographic problem, the scrambling map we
developed produces hard to optimize Ising instances. These instances function
as a filter for emerging combinatorial optimization accelerators that are
\emph{fundamentally} performing computation beyond local search.
 
\section{Code Availability}

Our implementations of the Stern algorithm and our parallel tempering solver
for the McEliece instances are publicly available as part of the PySA
Optimization Framework~\cite{pysa_stern}. The code to generate random instances
using the protocol presented in this paper is also available as part of
PySA~\cite{pysa_mceliece}.

\section{Acknowledgments}
S.M, H.M.B, and G.M are KBR employees working under the Prime Contract No.
80ARC020D0010 with the NASA Ames Research Center. All the authors acknowledge
funding from DARPA under NASA-DARPA SAA2-403688. We thank the NASA Advanced
Supercomputing (NAS) division for resources on the Pleiades cluster.  The
United States Government retains, and by accepting the article for publication,
the publisher acknowledges that the United States Government retains, a
nonexclusive, paid-up, irrevocable, worldwide license to publish or reproduce
the published form of this work, or allow others to do so, for United States
Government purposes.

\appendix
\section{Typical Kernel Size For Random Matrices}\label{appendix:kernel}

\begin{figure}
	\centering
	\includegraphics[width=0.9\textwidth]{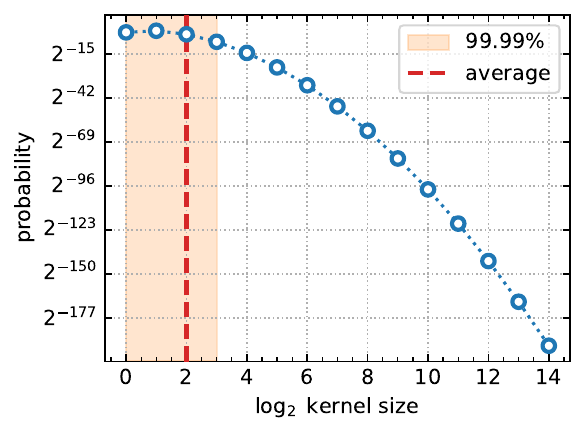}
  \caption{Distribution of the size of the kernel of random matrices extracted
  from $\mathbb{F}_2^{\,k\times k}$, in the limit of $k\to\infty$. The shaded
  area corresponds to the $99.99\%$ of the distribution, while the dashed line
  corresponds to its average.}
	\label{fig:kernel_size_distr}
\end{figure}

The probability $\mathcal{P}(\alpha)$ of a given random matrix
\mbox{$\mathbf{S}\in\mathbb{F}_2^{\,k\times k}$} having a rank ${\rm
rank}(\mathbf{S}) = k - \alpha$ in the limit of $k\to\infty$ can be expressed
as \cite{cooper2000distribution}:
\begin{equation}
	\mathcal{P}(\alpha) = 2^{-\alpha^2}\frac{\prod_{j=\alpha + 1}^\infty\left(1 - 2^{-j}\right)}
	{\prod_{j=1}^\alpha\left(1 - 2^{-j}\right)},
\end{equation}
where we used the convention that $\prod_{j=1}^0 (1 - 2^{-j}) = 1$. For
instance, the probability of randomly sampling a non-singular matrix is then
$\mathcal{P}(\alpha=0) \approx 0.288788$. As shown in
Fig.~\ref{fig:kernel_size_distr} and Table~\ref{tab:rank_distr},
$\mathcal{P}(\alpha)$ quickly goes to zero:

\FloatBarrier
\begin{table}[h!]
	\centering
	\begin{tabular}{|c|c|}
		\hline
		Matrix Rank [$k-\alpha$] & $\mathcal{P}(\alpha)$  \\
		\hline\hline
		0                        & 0.288788               \\
		1                        & 0.577576               \\
		2                        & 0.12835                \\
		3                        & $5.239 \cdot 10^{-3}$  \\
		4                        & $4.657 \cdot 10^{-5}$  \\
		5                        & $9.691 \cdot 10^{-8}$  \\
		6                        & $4.884 \cdot 10^{-11}$ \\
		\hline
	\end{tabular}
  \caption{Distribution of the rank or random matrices extracted from
  $\mathbb{F}_2^{\,k\times k}$, in the limit of $k\to\infty$.}
	\label{tab:rank_distr}
\end{table}
\FloatBarrier
\noindent with the number of elements in the kernel space having an average of
$\left\langle2^\alpha\right\rangle = 2$ and variance ${\rm Var}[2^\alpha] = 1$.

\section{Proof of the linear scrambling}

In this appendix we will compute the probability
$$
	\operatorname{Pr}\Big[E_{x}=\epsilon N, E_{y}=\epsilon N\Big]
$$
for the LSHWM to assign the energy value $\epsilon N$ simultaneously to two
states $x \neq y \in \{0,1\}^N$, for $\epsilon >0$. Clearly by linearity we
have that if $\epsilon > 0$ and either of these states is the zero vector, then
this probability must be zero. In the following we will assume that neither
$x$ nor $y$ are zero. Given a realization of disorder $\mathbf{S} = \{a_{ij}\}$ (\ie a
random matrix of the LSHWM ensemble), the energies of $x,y$ are
\begin{align*}
	E_x & = \sum_{j=1}^N \Big[ \sum_{k=1}^N a_{jk} x_k \pmod{2}\Big] 
    \equiv \sum_{j=1}^N E^{(j)}_x,  \\
	E_y & = \sum_{j=1}^N \Big[ \sum_{k=1}^N a_{jk} y_k \pmod{2}\Big]  
    \equiv \sum_{j=1}^N E^{(j)}_y.
\end{align*}
We partition the set $[N] \equiv \{1,2,\ldots,N\}$ as
\begin{align*}
	[N] =  & \{k \mid x_k = y_k = 0\} \cup \{k \mid x_k = y_k = 1\}            \\
	       & \cup \{k \mid x_k = 0, y_k = 1\} \cup \{k \mid x_k = 1, y_k = 0\} \\
	\equiv & (00) \cup (11) \cup (01) \cup (10)
\end{align*}
and coarse-grain the disordered degrees of freedoms $\{a_{jk}\}$ in $2^{\alpha
N}$ ``macrostates'' $\mathcal{S} \in \{0,1\}^{\alpha N}$ where each of the
$\alpha N$ coarse-grained binary variables $\mathcal{S}_{\gamma}^{(j)}$ of
$\mathcal{S}$ are functions of the matrix entries $\{a_{jk}\}_{jk}$. For each
$j=1,\ldots,N$ we define
$$
	\mathcal{S}_{\gamma}^{(j)} \equiv \sum_{k \in \gamma} a_{jk} \pmod{2}, 
    \quad \text{for $\gamma \in \Big\{(00),(01),(10),(11)\Big\}$}
$$
so that
$$
	\mathcal{S} = \Big(\; \Big\{ \mathcal{S}^{(j)}_{00} \Big\}_{j=1}^N, 
    \Big\{ \mathcal{S}^{(j)}_{01} \Big\}_{j=1}^N,
    \Big\{ \mathcal{S}^{(j)}_{10} \Big\}_{j=1}^N,
    \Big\{ \mathcal{S}^{(j)}_{11} \Big\}_{j=1}^N \;\Big).
$$
The integer $\alpha \in \{2,3,4\}$ will depend on the states $x,y$
(specifically, whether or not there exist indices $k$ where $x_k = y_k = 0$ so
that the set $(00)$ is nonempty, and analogously for $(01),(10)$ and $(11)$),
and we will distinguish cases. A specific choice of values for the disordered
variables $\{a_{jk}\}_{jk}$ will induce a unique macrostate, \ie an assignment
of Boolean values to all variables $\{\mathcal{S}_{\gamma}^{(j)}\}_{j,\gamma}$,
while a specific choice of values for
$\{\mathcal{S}_{\gamma}^{(j)}\}_{j,\gamma}$ will be compatible with multiple
``microstates'' $\{a_{jk}\}_{jk}$. Note that the variables $a_{ij}$ involved in
the definitions of the $\mathcal{S}_{\gamma}^{(j)}$ for different $\gamma$, are
all independent.

\begin{figure}
	\centering
	\includegraphics[width=0.9\textwidth]{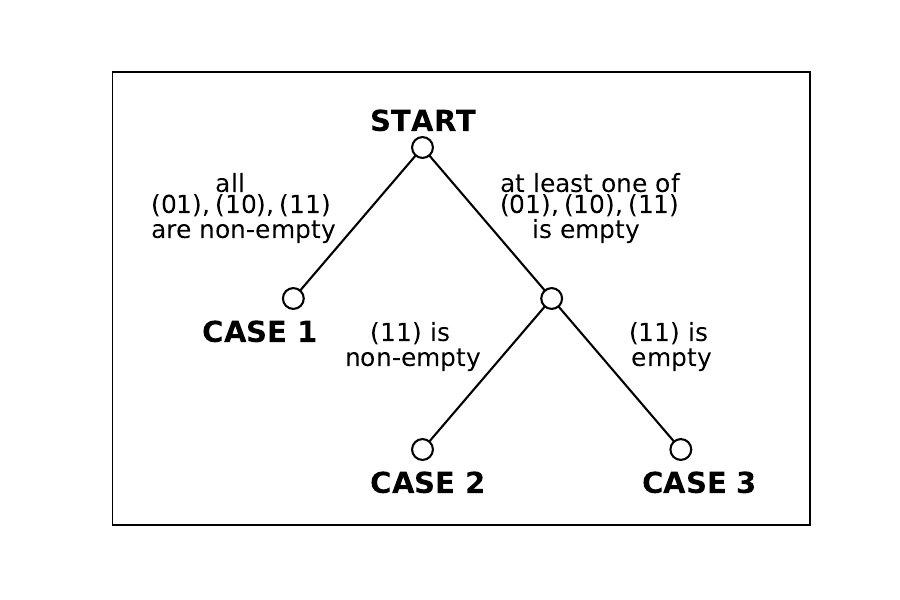}
  \caption{Logical flowchart of the three main cases for the calculation of the
  probability $\operatorname{Pr}\Big[E_{x}=\epsilon N, E_{y}=\epsilon N\Big]$ in the LSHWM.}
	\label{fig:cases_flowchart}
\end{figure}

Then the energies of the states $x,y$ we can be written as a function of the
macrostate variables:
\begin{align*}
	E_x & = \sum_{j=1}^N E_x^{(j)} = \sum_{j=1}^N 
    \Big[ \mathcal{S}_{11}^{(j)} + \mathcal{S}_{10}^{(j)} \pmod{2}\Big] \\
	E_y & = \sum_{j=1}^N E_y^{(j)} = \sum_{j=1}^N 
    \Big[ \mathcal{S}_{11}^{(j)} + \mathcal{S}_{01}^{(j)} \pmod{2}\Big]
\end{align*}
Observe in particular that the energies do not depend on the macrostate
variables $\mathcal{S}_{00}^{(j)}$. In order to compute the probability of
having $E_x=E_y=\epsilon N$ we need to count how many ways we can choose values
for the macrostate variables that will give us this energy value for both $E_x$
and $E_y$, and then weight each choice by its appropriate probability induced
by the distribution of the microscopic degrees of freedom $\{a_{jk}\}_{jk}$
compatible with that choice. We distinguish the following three main cases,
summarized in Fig.~\ref{fig:cases_flowchart}.

\begin{itemize}
  \item \textbf{Case 1: the three subsets $\gamma = (11),(01),(10)$ are
    non-empty.}
        Let us fix a vector $\mathcal{S}_{11} = (\mathcal{S}_{11}^{(1)},
        \ldots,\mathcal{S}_{11}^{(N)}) \in \{0,1\}^N$. In order to have $E_x =
        \epsilon N$ we need to have $\epsilon N$ indices $0 \leq j \leq N$
        where $E^{(j)}_x = 1$. This fixes the values of
        $\mathcal{S}^{(j)}_{10}$ since we need to have $\epsilon N$ indices
        where $\mathcal{S}^{(j)}_{10} = 1 - \mathcal{S}^{(j)}_{11}$ and the
        other $(1-\epsilon)N$ must be $\mathcal{S}^{(j)}_{11} =
        \mathcal{S}^{(j)}_{10}$.  Analogously we need to choose $\epsilon N$
        values of $j$ where $E_y^{(j)}=1$. These can be chosen independently
        from the choices of $E_x^{(j)}$. This fixes the values
        $\mathcal{S}^{(j)}_{10}$.
	      \begin{itemize}
		      \item \textbf{Case 1.A: the subset $(00)$ is non-empty}
                Then $\alpha=4$ and we have the $N$ degrees of freedom in
                $\mathcal{S}^{(j)}_{00}$ that are completely free and do not
                affect the energies at all. So for a fixed $\mathcal{S}_{11} =
                (\mathcal{S}_{11}^{(1)}, \ldots,\mathcal{S}_{11}^{(N)}) \in
                \{0,1\}^N$ we have
		            $$
			            2^{N}\binom{N}{\epsilon N}^2
		            $$
                choices to achieve that $E_x=E_y=\epsilon N$. Which means that
                at the level of the macrostates $\mathcal{S}$, the number $n_M$
                of choices that give $E_x=E_y=\epsilon N$ is
		            $$
			            n_M = 2^{2N}\binom{N}{\epsilon N}^2
		            $$
                What is left to do is to compute the entropies of the
                macrostates. A given value $\mathcal{S}_{\gamma}^{(j)}$ is
                compatible with half of the microstates of the variables
                $a_{ij}$ that appear in its definition, and all of these
                choices are independent for different $j$ and $\gamma$, so we
                get (remember that here $\alpha=4$)
		            $$
			            2^{-\alpha N}\Big[ 2^N \Big]^N 2^{2N}\binom{N}{\epsilon N}^2
                    = 2^{N^2} 2^{-2N}\binom{N}{\epsilon N}^2
		            $$
                out of a total of $2^{N^2}$ microstates, so the probability we
                seek is given by
		            \begin{align*}
			            \operatorname{Pr}\Big[E_{\sigma}=\epsilon N, 
                    E_{\tau}=\epsilon N\Big] = 2^{-2N}
                      \binom{N}{\epsilon N}^2 \equiv P_\epsilon
		            \end{align*}
		      \item \textbf{Case 1.B: the subset $(00)$ is empty}
                In this case $\alpha=3$ and we have already fixed all
                macrostate variables. By an analogous calculation to the Case
                1.A we have
		            $$
			            n_M = 2^{N}\binom{N}{\epsilon N}^2
		            $$
                choices for the macrostate variables compatible with
                $E_x=E_y=\epsilon N$. In the language of the microstates, we
                have
		            $$
			            2^{-\alpha N}\Big[ 2^N \Big]^N 2^{N}\binom{N}{\epsilon N}^2
                    = 2^{N^2} 2^{-2N}\binom{N}{\epsilon N}^2
		            $$
                choices out of $2^{N^2}$ total microstates. This gives the same
                value $P_{\epsilon}$ for the probability, as before.
	      \end{itemize}
  \item \textbf{Case 2: $(11)$ is non-empty, and at least one between
    $(01),(10)$ is empty.}
        Notice that in this case we must have that \emph{exactly} one between
        $(01),(10)$ is empty (if both are empty then $x=y$, a possibility that
        we have ruled out by assumption). Since the states $x,y$ play a
        perfectly symmetrical role in the probability we are trying to compute
        we can assume without loss of generality that $(01)$ is non-empty and
        $(10)$ is empty. In this case, in order to have $E_x = \epsilon N$ we
        have to choose $\epsilon N$ variables $\mathcal{S}_{11}^{(j)}$ to be
        set to one, and all the others to zero ($\binom{N}{\epsilon N}$
        choices). This fixes the vector $\mathcal{S}_{11}$. In order to obtain
        $E_y = \epsilon N$ we have to choose $\epsilon N$ variables
        $\mathcal{S}_{01}^{(j)}$ to be set equal to $\mathcal{S}_{01}^{(j)} =
        1-\mathcal{S}_{11}^{(j)}$ and the others to $\mathcal{S}_{01}^{(j)} =
        \mathcal{S}_{11}^{(j)}$ (another $\binom{N}{\epsilon N}$ choices). This
        fixes the vector $\mathcal{S}_{01}$.
	      \begin{itemize}
		      \item \textbf{Case 2.A: the subset $(00)$ is non-empty}
                Then $\alpha=3$ and the proof follows Case 1.A with $n_M = 2^N
                \binom{N}{\epsilon N}$ which leads to a probability of
                $P_\epsilon$ as before.
		      \item \textbf{Case 2.B: the subset $(00)$ is empty}
                Then $\alpha=2$ and the proof follows Case 2.A with $n_M =
                \binom{N}{\epsilon N}$ which leads to a probability of
                $P_\epsilon$ as before.
	      \end{itemize}

	\item \textbf{Case 3: $(11)$ is empty.}
        In this case we must have that both $(01)$ and $(10)$ are non-empty,
        otherwise at least one between $x$ and $y$ is the zero vector, a
        possibility that we have ruled out by assumption. In order to have $E_x
        = E_y = \epsilon N$ we have to set to one exactly $\epsilon N$
        variables $\mathcal{S}_{01}^{(j)}$, and exactly $\epsilon N$ variables
        $\mathcal{S}_{10}^{(j)}$, independently. This gives $\binom{N}{\epsilon
        N}^2$ choices. This fixes the vectors $\mathcal{S}_{01}$ and
        $\mathcal{S}_{10}$.
	      \begin{itemize}
		      \item \textbf{Case 3.A: the subset $(11)$ is non-empty}
                Then $\alpha=3$ and the proof follows Case 1.A with $n_M = 2^N
                \binom{N}{\epsilon N}$ which leads to a probability of
                $P_\epsilon$ as before.
		      \item \textbf{Case 3.B: the subset $(11)$ is empty}
                Then $\alpha=2$ and the proof follows Case 2.A with $n_M =
                \binom{N}{\epsilon N}$ which leads to a probability of
                $P_\epsilon$ as before.
	      \end{itemize}
\end{itemize}

\section{Computational Algorithms}\label{appendix:isd}

\emph{Stern's Algorithm}. Let us give a detailed overview of the information
set decoding (ISD) strategy. Given a binary matrix $H\in \FF_2^{(N-k)\times n}$
and a vector $y\in \FF_2^{(N-k)}$, Stern's algorithm searches for a vector
$x\in \FF_2^n$ of Hamming weight $\norm{x}\leq t$ such that
\begin{equation}
     H x \equiv y \mod 2\,.
\end{equation}
Alternatively, given a binary matrix $H\in \FF_2^{(N-k)\times n}$, search for a
non-zero vector $x\in \FF_2^N$ of minimum Hamming weight such that
\begin{equation}
     H x \equiv 0 \mod 2\,.
\end{equation}
The steps involved in information decoding are as follows:
\begin{enumerate}
    \item (Initialization) Initialize the augmented matrix
          \begin{equation}
              A \equiv (H | y)
          \end{equation}
          where the $N$ columns of $H$ are the ``principal'' columns and the
          additional column $y$ is the sole augmented column. If calculating
          code distance, then we simply have $y=0$, so augmentation is not
          necessary.
    \item (IR Partition) Randomly partition the $N$ principal columns of $A$
          into an information/redundancy set pair $(I,R)$. Perform binary
          Gauss-Jordan elimination on $A$ over the columns indexed by $R$,
          restarting with a new random pair if the submatrix $A_{|R}$ is not
          full rank.  Denote by $y'$ the final column of $A$ after elimination.
    \item (Combinatorial Search) Search for vectors $x_0\in \FF_2^{N-k}$ and
          $x_1\in \FF_2^k$ such that the objective
          \begin{equation}\label{eq:stern1}
              \mathcal{L}(x_0, x_1) =  \norm{y' - (x_0 + A_{|I} x_1)} \,,
          \end{equation}
          is minimized to 0 subject to the constraint $\norm{x_0} + \norm{x_1}
          \leq t$.  If calculating the code distance, then perform the
          minimization subject to the constraint that $\norm{x_0} >0$.  The
          solution $x$ is gathered from $x_0$ and $x_1$ so that $x_{|R} = x_0$
          and $x_{|I}=x_1$.
    \item (Result) If minimization succeeds, return the solution $x$.
          Otherwise, return to step 2 until termination/time-out.
\end{enumerate}

\begin{algorithm}[t]
  \caption{Stern's Algorithm, ``heavy'' version. The subroutine
           $\mathrm{GJE}(A, R)$ performs Gauss-Jordan elimination on the matrix
           $A$ using the list of columns $R$ as pivots.}\label{alg:stern}
  \begin{algorithmic}[1]
  \Procedure{Stern}{$H\in\mathbb{F}_2^{(N-k)\times n}$, $y\in F_2^{(N-k)}$}
  \State Initialize $A \gets (H|y)$, which has $N$ principal columns and $y$ is the sole augmented column.
  \Repeat
  \State Randomly sample an IR partition $(I, R)$ of the principal columns of $A$.
  \State $A'= (H'|y')\gets\mathrm{GJE}(A, R)$
  \Until $A'_{|R}$ is full rank.
  \State Randomly split $I$ into $I_1$ and $I_2$.
  \For{Each combination ($i_1$,\ldots, $i_p$) from $I_1$}
  \For{Each combination ($j_1$,\ldots, $j_p$) from $I_2$}
  \State Let $b\gets \sum_{n}A'_{i_n} + \sum_m A'_{j_m} +y'$
  \If {$\mathrm{HW}(b) = t-2p$}
  \State $\mathcal{E}\gets \{k| b_k>0\}\cup \{i_1,\ldots, i_p\}\cup \{j_1,\ldots, j_p\}$
  \State \textbf{Return} the error set $\mathcal{E}$
  \EndIf
  \EndFor
  \EndFor
  \State \textbf{Return} failure.
  \EndProcedure
  \end{algorithmic}
\end{algorithm}

Stern's algorithm is an implementation of ISD where the combinatorial search
step randomly splits the information set $I$ into two sets, $I_1$ and $I_2$
then searches for a linear combination consisting of $y'$ with $p$ columns from
$I_1$ and $p$ columns from $I_2$ that has a Hamming weight of exactly $w-2p$.
If such a combination is found, the indices of the non-zero bits are precisely
the error bits in the redundancy set, and the $2p$ columns are the remaining
error bits contained in the information set.  The specific procedure is shown
in Algorithm~\ref{alg:stern}. Our implementation is a ``heavy'' version of
Stern's algorithm that skips collision checking on the $2p$ columns before
calculating the full Hamming weight. The collision check was observed to have a
negative impact for all parameter sets in our benchmarks. The $p$ integer
parameter (unrelated to $p$ as in $p$-local Hamiltonian) of Stern's algorithm
was fixed to $p=1$, as larger values did not benefit the theoretical success
probability for the parameter sets used here.

When mapped to a corresponding $p$-local Hamiltonian, the IR partition defines a gauge
transformation on the spin variables such that $N-k$ terms are reduced to
linear interactions.  This transformation simplifies the chosen terms in the
Hamiltonian in exchange for potentially higher-order interactions on the
remaining terms. ISD is therefore well-suited for an unstructured $p$-local
Hamiltonian where $p$ is already large and there is no clear interaction
locality.  On the other hand, if $p$ is small, local search heuristics may be
more effective, while ISD unnecessarily transforms the problem to include
higher-order interactions.

\emph{Parallel Tempering}.
The parallel tempering algorithm, also known as replica exchange Monte Carlo,
is a Markov Chain Monte Carlo algorithm that performs Metropolis-Hastings
sampling on multiple replicas at different temperatures
\cite{swendsen1986replica, marinari1992simulated, hukushima1996exchange}. The
replicas converge to thermal equilibrium by including a move to swap two
replicas in neighboring temperatures and occasionally performing this move
under the Metropolis-Hastings criterion.  This accelerates the convergence of
low-temperature states of Ising spin glasses compared to sampling at a single
temperature alone. A variation of PT that performs energy-conserving
(``isoenergetic'') cluster moves across replicas (called PT-ICM) has been
especially successful in benchmarks against quantum annealing for finding the
ground-state energy of finite-dimensional Ising spin glasses
\cite{zhu2015efficient,mandra2016strengths,mandra2017pitfalls}. As the
instances we generate are dense, rather than finite-dimensional, there is no
computational advantage in performing cluster moves, so we only benchmark the
ordinary PT algorithm rather than PT-ICM.  The procedure of our implementation
is given in Algorithm~\ref{alg:pt}.

\begin{algorithm}[t]
  \caption{Parallel tempering algorithm for a $p$-local Ising Hamiltonian.
           $\mathrm{Rand}()$ is a call to a uniformly random floating point
           point number in the interval $[0, 1)$.}\label{alg:pt}
  \begin{algorithmic}[1]
  \Procedure{PT}{Hamiltonian $H(\sigma)$ for an $[N, k, t]$ code, $N_T$, $\beta_i$, $N_{\mathrm{steps}}$}
  \State Randomly initialize replicas $\sigma^{(i)},i=1,\ldots N_T$ .
  \For{$t \in {1,\ldots, N_{\mathrm{steps}}}$}
  \For{$i\in\{1,\ldots N_T \}$}
      \For{$j\in {1,\ldots k}$} \Comment{Sweep replica $i$}
      \State Calculate $\Delta E$ of flipping spin $j$ in  $H(\sigma_i)$
      \State $p\gets e^{-\beta\Delta E}$
      \If{$\Delta E < 0$ or $\mathrm{Rand}() < p$ }
          \State $\sigma_{j}^{(i)} \gets -\sigma_{j}^{(i)}$
      \EndIf
      \If{$H(\sigma^{(i)}) = t - N$}
      \State $x_j \gets  (1-\sigma^{(i)}_j)/2$
      \State \textbf{Return} decoded message $x$.
      \EndIf
      \EndFor
  \EndFor
  \For{$i\in\{1,\ldots N_T-1\}$} \Comment{Replica exchange moves}
  \State $f \gets (\beta_{i+1} - \beta_{i})(E_{i+1} - E_{i})$
  \If {$f \geq 0$ or $\mathrm{Rand}() < e^{f}$}
  \State Swap replicas $i$ and $i+1$
  \EndIf
  \EndFor
  \EndFor
  \State \textbf{Return} failure.
  \EndProcedure
  \end{algorithmic}
\end{algorithm}

\begin{table}[]
    \centering
    \begin{tabular}{cccc}
         \hline
         $N_T$ & $\beta_{\mathrm{min}}$  &  $\beta_{\mathrm{max}}$ &   $N_{\mathrm{Steps}}$ \\
         \hline
         16  & 0.1 & 1.0 & $10^7$\\
    \end{tabular}
    \caption{Table of parameters used for parallel tempering.}
    \label{tab:ptparams}
\end{table}

We used a geometrically spaced array of inverse temperatures, \ie $\beta_i =
e^{x \beta_{\mathrm{min}} + (1-x_i)\beta_{\mathrm{max}}}$ where $x_i = (N_T -i
)/(N_T - 1)$ where $i=1,2,\ldots N_T$ is the index of each replica temperature,
$\beta_{\mathrm{min}}$ is the smallest inverse temperature, and
$\beta_{\mathrm{max}}$ is the largest inverse temperature.  We chose $N_T$,
$\beta_{min}$ and $\beta_{max}$ so that the time-to-solution for a few
instances at $k=16$ through $20$ was satisfactory across individual runs on a
personal computer and the number of accepted replica exchange moves was not
depressed at any temperature. We did not attempt additional optimization of the
array of temperatures. Because the scrambled energy landscape thwarts any
algorithm relying on local moves, it is unlikely that the scaling is sensitive
to any further optimization of the PT parameters. The parameters of our PT
benchmarks are fully specified in Table~\ref{tab:ptparams}.

\section{Generation of McEliece Instances}\label{appendix:generator}

We used random Goppa codes as the underlying code for the McEliece protocol,
but allowed the ability to tune all code parameters to generate computationally
tractable benchmarking instances to perform our scaling analysis.  We
implemented a random instance generator with the McEliece cryptographic
protocol which takes the desired code length $N$ and error correcting distance
$t$ as inputs, as well as an additional parameter $m\geq\lfloor\log_2 N\rfloor$
for the Goppa code.  The message length is determined as $k=N - tm$.  Instance
generation is performed in following steps:
\begin{enumerate}
  \item Generate a random $k$-bit plain-text message $q$ to encode and a random
        weight $t$ error $\epsilon$. 
  \item Generate a random irreducible degree $t$ polynomial $P$ over
        $\mathbb{F}_{2^m}$. This can be done by rejection sampling on random
        polynomials over $\mathbb{F}_{2^m}$. The expected number of trials of
        rejection sampling before finding an irreducible polynomial is no
        greater than the degree of the polynomial, \ie $O(t)$, while the
        average cost of testing for irreducibility via Rabin's algorithm is
        $O(m t^2)$ (the polynomial degree times the cost of polynomial
        multiplication) \cite{panarioAnalysis98}. Thus, the average case
        complexity of generating a random irreducible polynomial is $O(m t^3)$.
  \item Generate a random Goppa code by randomly sampling $N$ elements
        $\alpha_i$ from the finite field $\mathbb{F}_{2^m}$. The generator
        matrix $\mathbf{G}$ and error correcting algorithm are defined using
        the value of $P$ evaluated on each of the $N$ elements (see,
        \emph{e.g.},~\cite{pattersonAlgebraic75, bernstein2008attacking}). For
        consistency with the desired message length, the Goppa code is rejected
        and re-sampled from step 2 if the null space of $\mathbf{G}$ is larger
        than $k$.
  \item Randomly generate $\mathbf{S}$ and $\mathbf{P}$ to construct the public
        key $\mathbf{G}'$ from $\mathbf{G}$. 
  \item Encrypt the message via Eq.~\eqref{eq:mcelieceEncrypt}. 
\end{enumerate}
Thus, we have the $\mathbf{G'}$ and $q'$ which completely define the McEliece
problem instance and hence the scrambled Ising Hamiltonian. The instance is
solved if the plain-text message $q$ (or equivalently the error vector
$\epsilon$) is found.  

While we have made every effort to test the correctness of the generator,
encoder, and decoder, our aim was to generate cryptographic problems with
tunable attack complexity and thus made no attempt at ensuring compliance with
the Classic McEliece specifications.  One important difference is that our
time-complexity to generate a public/private key pair is dominated by the
second step--finding a random irreducible polynomial over $\mathbf{F}_{2^m}$ as
$O(m t^3)$. The Classic McEliece specification does not use Rabin's algorithm
for the secure generation of a random irreducible polynomial since its success
rate decreases with the desired security of the public/private key pair. 

\bibliography{refs}
\bibliographystyle{apsrev4-2}

\end{document}